\documentstyle[epsf]{mn}
\onecolumn

\newcommand{\subI}{_{\rm _I}}
\newcommand{\subR}{_{\rm _R}}
\newcommand{\subs}[1]{{\rm _{s#1}}}
\newcommand{\sube}[1]{{\rm _{e#1}}}
\newcommand{\subi}[1]{{\rm _{i#1}}}
\newcommand{\dR}{\delta\subR}
\newcommand{\dI}{\delta\subI}
\newcommand{\dIO}{\delta_{_ {\rm IO}} }
\newcommand{\dC}{\delta_{_ {\rm C}} }
\newcommand{\dQ}{\delta_{_ {\rm Q}} }
\newcommand{\vff}{v_{\rm ff}}
\newcommand{\rhoa}{\rho_{\rm a}}

\newcommand{\vb}{{\bf v}}

\newcommand{\tG}{{\tilde \Gamma}}
\newcommand{\tL}{{\tilde \Lambda}}
\newcommand{\psic}{\psi_{\rm c}}
\newcommand{\psiei}{\psi_{\rm ei}}

\newcommand{\kb}{k_{_{\rm B}}}

\newcommand{\GS}{\ga}
\newcommand{\LS}{\la}
\newcommand{\etal}{et al.\ }
\newcommand{\eg}{e.g.\ }

\title{Perturbative Analysis of Two-Temperature Radiative Shocks
with Multiple Cooling Processes}
\author[C.~J.~Saxton \& K.~Wu]{
Curtis~J.~Saxton$^{1,2}$ \& Kinwah~Wu$^1$\\
$^1$ Research Centre for Theoretical Astrophysics, School of Physics, 
University~of~Sydney, NSW 2006, Australia \\
$^2$ Sir~Frank~Packer Department of Theoretical Physics, School of Physics, 
University~of~Sydney, NSW 2006, Australia
}

\date{Received: }

\begin{document}

\maketitle

\begin{abstract} 
The structure of the hot downstream region below a radiative accretion 
shock, such as that of an accreting compact object, may oscillate due 
to a global thermal instability. The oscillatory behaviour depends on 
the functional forms of the cooling processes, the energy exchanges of 
electrons and ions in the shock-heated matter, and the boundary 
conditions. We analyse the stability of a shock with unequal electron 
and ion temperatures, where the cooling consists of thermal 
bremsstrahlung radiation which promotes instability,
plus a competing process which tends to stabilize the shock. The effect 
of transverse perturbations is considered also. As an illustration, we 
study the special case in which the stabilizing cooling process is of 
order 3/20 in density and 5/2 in temperature, which is an approximation 
for the effects of cyclotron cooling in magnetic cataclysmic variables.
We vary the efficiency of the second cooling process, the strength of 
the electron-ion exchange and the ratio of electron and ion pressures 
at the shock, to examine particular effects on the stability properties
and frequencies of oscillation modes.

\end{abstract}


\section{Introduction} 

Depending on the form of cooling processes present, radiative shocks 
may suffer thermal instabilities. The resulting sequence of oscillatory 
modes has frequencies and stability properties determined by the 
functional form of the energy loss terms, the rate of energy exchange 
between ions and electrons, and the particular boundary conditions of 
the system. Oscillations modulating the structure of the hot downstream 
column cause variations in the observable emissions. Radiative shocks 
in several different astrophysical systems show thermal instability, for 
example the interactions between supernova shocks and the interstellar 
medium, and accretion streams onto magnetic and non-magnetic compact 
objects. Many investigations have considered the stability properties 
of radiative shocks, either through numerical calculations or 
perturbative analyses (\eg Falle 1975, 1981;
Langer, Chanmugam \& Shaviv 1981, 1982; 
Chevalier \&  Imamura 1982; Imamura, Wolff \& Durisen 1984; 
Chanmugam, Langer \& Shaviv 1985; Imamura 1985; Bertschinger 1986; 
Innes, Giddings \& Falle 1987a, b; Gaetz, Edgar \& Chevalier 1988; 
Wolff, Gardner \& Wood 1989; Imamura \& Wolff 1990; 
Houck \& Chevalier 1992; Wu, Chanmugam \& Shaviv 1992; 
T\'{o}th \& Draine 1993; Dgani \& Soker 1994; Imamura \etal\ 1996;
Saxton, Wu \& Pongracic 1997; Saxton \etal 1998).

In accreting white dwarfs the flow meets a stand-off accretion shock
where the inwards falling matter is suddenly decelerated to subsonic 
settling flow. The shock sits above the white dwarf surface at a 
height determined by the path length of the radiative cooling of the 
downstream material $x\subs{}\approx{\frac14}\vff t_{\rm cool}$, which 
depends on the free-fall velocity of the white dwarf
$\vff={(2GM_{\rm wd}/R_{\rm wd})}^{1/2}$, and on the cooling timescale
$t_{\rm cool}\sim \kb T/\Lambda$, where $\Lambda$ is a radiative cooling 
loss function. In cataclysmic variables (see Warner 1995),
the white dwarf accretes mass from a close red dwarf companion star.
Thermal bremsstrahlung cooling is an important cooling process
in the post-shock region. In cases where the magnetic field of the 
white dwarf is strong, ($B_{\rm wd}\GS 10$~MG), 
(\eg Lamb \& Masters 1979; King \& Lasota 1979), cyclotron radiation 
provides another important cooling process. These competing processes 
have different dependences on the local flow variables, and therefore 
they influence the shock stability and oscillatory behaviour differently.

Langer \etal\shortcite{langer81} carried out time-dependent numerical 
investigations of accretion onto a white dwarf in a planar geometry.
Their results show quasi-periodic oscillations
due to thermal instability when the cooling is via bremsstrahlung 
radiation. In the linear perturbative analysis as in
Chevalier \& Imamura \shortcite{chevalier}, 
the radiative shock with bremsstrahlung cooling
was shown to have a stable fundamental mode and unstable overtones.
They also investigated other cases with a power-law cooling function
in density $\rho$ and temperature $T$,
$\Lambda \propto \rho^a T^b$, 
with various choices of the power indices
(\eg $(a,b)=(0.5,2)$ for bremsstrahlung cooling).
These investigations show that radiative shocks where the index $b$
is higher are more strongly stabilized against perturbations.

Bertschinger \shortcite{bertschinger} studied the effect of placing the 
radiative shock in a spherical geometry. Again the cooling function was 
a single power-law of density and temperature.
In addition to radial perturbations,
transverse perturbations were investigated,
which are expressed in terms of a scaled transverse wavenumber
in addition to the usual oscillation frequency.
The modes that are stable in the purely radial analysis of
Chevalier \& Imamura (1982)
are destabilized in the presence of transverse oscillation,
but all modes are stable in the limit of large wavenumber.

The stability of physically extended shocks was studied by
Houck \& Chevalier \shortcite{houck}  for radial accretion onto compact 
objects, with a Newtonian gravitational potential.
The adiabatic index of the gas
and the indices of the single power-law cooling term were varied,
In the limit of a non-extended shock envelope
their calculations yield the same results as
Chevalier \& Imamura \shortcite{chevalier}. 
However even a slight spherical extension of the post-shock region
has an effect on the stability: stabilizing the first and second harmonics 
but destabilizing the higher order modes.

Physical factors other than the cooling processes and geometry
also affect the stability properties of a radiative shock.
The effects of mass loss from the sides of the post-shock accretion column
were examined by Dgani and Soker \shortcite{dgani} 
by including a sink term in the mass continuity equation.
They found that the transverse leakage has a stabilizing effect
in the presence of several single power-law radiative cooling processes,
and this stabilization was less effective
when the cooling had a lower temperature index.
T\'{o}th \& Draine \shortcite{toth} 
investigated radiative shocks that
were subject to thermal instability
but were supported by a transverse magnetic field.
This magnetic pressure support
is decisive in determining the stability properties,
rather than the characteristics of the radiative cooling alone.

The consequences of inequality between electron and ion temperatures,
and the effect of the energy exchange between these fluid components,
was considered by Imamura \etal \shortcite{aboasha}, 
with the effects of bremsstrahlung and Compton cooling.
Their perturbations included a transverse component,
as in Bertschinger \shortcite{bertschinger},
although the geometry of the system was planar.
Modes with transverse components were less stable than
purely longitudinal equivalents,
and the presence of the electron-ion exchange also promoted instability.
These new effects were qualitatively similar under different choices
of the temperature index of the single cooling process.

In these studies, the usual treatment of the radiative cooling processes
was to use an approximate energy loss term that is a single power-law in 
local fluid properties. In the presence of multiple cooling processes,
intermediate power-law indices tend to be used. This is not an adequate 
treatment for the detailed effects of the competition between cooling 
processes, which is especially important when one process, such as 
thermal bremsstrahlung radiation, promotes thermal instability, while 
another process, such as cyclotron emission, has a stabilizing influence.
The complicated competition between such processes must be treated
by taking an explicit sum of independent power-law contributions
for the respective processes.

Chanmugam \etal \shortcite{cls} 
performed numerical calculations of the accretion onto a magnetic white 
dwarf with a sum of separate bremsstrahlung and cyclotron power-law 
contributions. When cyclotron cooling was strong,
the shock was stabilized against oscillations.
Further examinations (Wu \etal 1996) revealed that the stabilizing 
influence depends on the magnetic field strength.
In a weak-field regime ($B\LS10$MG)
the competition of the bremsstrahlung and cyclotron cooling,
each dominating in different phases of a two-phase oscillatory cycle,
allows the perpetuation of small-amplitude oscillations
(Wu \etal 1992).  

Hujeirat \& Papaloizou \shortcite{hujeirat}
performed one- and two-dimensional numerical calculations to 
investigate the structure and time-dependent behaviour of accretion 
onto compact objects in the presence of bremsstrahlung cooling, and 
with consistent solution of the radiative transport and MHD equations.
Magnetic field strengths were investigated up to
$\sim10^{-2}$ times the values found in
magnetic cataclysmic variables (mCVs).
They found that radiation-matter coupling in the flow below
the photosphere reduces the amplitudes of the shock oscillations,
but that the oscillations are sensitive to the lower boundary condition.
In weak-field cases, latitudinal flows out of the accretion column
reduce or suppress oscillations;
and in stronger field cases there remains some transverse kinks
in the magnetic field near the shock,
which tends to suppress oscillations because of the effect found by
T\'{o}th \& Draine \shortcite{toth}.

A composite cooling function was used in
the single-temperature perturbative stability analyses of
Saxton \etal \shortcite{saxton97} 
and Saxton \etal \shortcite{saxton98}.
The energy loss term consisted of a sum of terms for
thermal bremsstrahlung radiation 
($\Lambda_{\rm br}\propto\rho^2 T^{0.5}$) 
and a second power-law process with a destabilizing influence
($\Lambda_{_2}\propto\rho^a T^b$ for $b\GS1$).
The cases considered included that of an effective cyclotron cooling term 
($\Lambda_{\rm cy}\propto\rho^{0.15} T^{2.5}$)
appropriate for the conditions and flow geometry
of accreting magnetic white dwarfs.
The analyses examined effects of different values of
the efficiency of the second cooling process,
which depends on the magnetic field strength
in the cyclotron-cooling case.
Increasing the cyclotron cooling strength stabilized each mode 
monotonically, but this happened in a more fundamentally complicated manner 
than straightforward comparison of cooling and oscillatory timescales would 
suggest. The order and manner in which the harmonics stabilized depends on
the indices of the second cooling process.

In the paper we generalize the previous work in
Saxton \etal (1997) and Saxton \etal (1998)
for bremsstrahlung-cyclotron radiative accretion shocks in mCVs,
by including the effects of unequal electron and ion temperatures
and the possibility of transverse perturbations (for a corrugated shock).
The equations for a radiative post-shock accretion flow are presented in
section\ \ref{'formulation'}
and the terms and formulation of the perturbation analysis are developed 
in section\ \ref{'perturbation'}.
In section\ \ref{'results'} we calculate and interpret
the frequency and stability properties of the eigenmodes,
and explore the effects of varying
the parameters of the two-temperature condition,
and the relative strengths of the competing cooling processes.
In section\ \ref{'discussion'}
we discuss the significance of our results
in contrast with previous studies dealing with
cases with different physical processes
or which were less general.
We conclude in section\ \ref{'conclusions'}.

\section{Formulation}
\label{'formulation'}

We assume the adiabatic index $\gamma={5/3}$ for an ideal 
gas, and the equation of state
\begin{equation}
P={{\rho\kb T}\over{\mu m_{_{\rm H}}}},
\end{equation}
where
$\kb$ is the Boltzmann constant,
and $m_{_{\rm H}}$ is the mass of the a hydrogen atom.
The time-dependent mass continuity, momentum and energy equations for
the planar post-shock accretion flow are:
\begin{equation}
\left({
{\partial\over{\partial t}} + \vb\cdot\nabla
}\right)
\rho
+\rho\left({\nabla\cdot\vb}\right)
= 0,
\end{equation}
\begin{equation}
\rho\left({
{\partial\over{\partial t}} + \vb\cdot\nabla
}\right)
\vb
+\nabla P
= 0,
\end{equation}
\begin{equation}
\left({
{\partial\over{\partial t}}
+\vb\cdot\nabla
}\right) P
-\gamma{P\over\rho}
\left({
{\partial\over{\partial t}}
+\vb\cdot\nabla
}\right) \rho
= -(\gamma-1)\Lambda,
\end{equation}
\begin{equation}
\left({
{\partial\over{\partial t}}
+\vb\cdot\nabla
}\right) P\sube{}
-\gamma{{P\sube{}}\over\rho}
\left({
{\partial\over{\partial t}}
+\vb\cdot\nabla
}\right) \rho
= (\gamma-1)(\Gamma-\Lambda),
\end{equation}
where $P$, $P\sube{}$, $\vb$ and $\rho$
are respectively the total pressure, electron pressure, velocity and 
density of the stream;
$\Lambda$ is the composite radiative cooling function,
and $\Gamma$ is the electron-ion energy exchange.
\begin{equation}
\Gamma={{4\sqrt{2\pi}e^4n\sube{}n\subi{}\ln C}\over{m\sube{} c}}
\left[{
{\theta\subi{}-(m\sube{}/m\subi{})\theta\sube{}}
\over{(\theta\sube{}+\theta\subi{})^{3/2}}
}\right]
\end{equation}
where
$\ln C$ is the Coulomb logarithm,
$c$ is speed of light,
$e$ is the electron charge,
$\theta\sube{}=\kb T\sube{}/m\sube{} c^2$ and
$\theta\subi{}=\kb T\subi{}/m\subi{} c^2$
are the electron and ion temperatures
(see, e.g. Melrose 1986).
The electron and ion number densities
are $n\sube{}=Z\rho/(Zm\sube{}+m\subi{})$
and $n\subi{}=\rho/(Zm\sube{}+m\subi{})$
if the ion charge is $Z$ ($Z=1$ for completely ionized hydrogen plasma).

Optically thin thermal bremsstrahlung radiation
provides one of the two cooling processes present,
and the other is taken to be a single power-law cooling term.
To simplify the analysis, we express the total cooling function
in terms of the bremsstrahlung cooling term
and a function expressing its functional form compared with 
the bremsstrahlung cooling term. The parameter $\epsilon\subs{}$ is
the efficiency of the second cooling process compared to bremsstrahlung 
cooling, evaluated at the shock, which depends non-linearly on
the white dwarf's magnetic field strength, mass and radius,
and the pre-shock density and cross-sectional circumference and area of
the accretion flow.
\begin{equation}
\Lambda_{_T} \equiv \Lambda_{_{\rm br}}+\Lambda_{_2}
=\Lambda_{\rm br}
\left[
1+\epsilon\subs{}\left({P\sube{} \over{P\sube{,s}}}\right)^\alpha
\left({\rho\over{\rho_{\rm s}}}\right)^{-\beta}
\right].
\label{'eq.lambda.total'}
\end{equation}
where $\rho\subs{}$ and $P\sube{,s}$
are the density and electron pressure at the shock,
$\Lambda_{\rm br}={\mathcal C}\rho^2\left({P\sube{}/\rho}\right)^{1/2}$
and the constant
${\mathcal C} = (2\pi k_{\rm _B}/3 m_e)^{1/2} (2^5\pi e^6/3 h m_e c^3) 
 (\mu/ k_{\rm _B} m_p^3)^{1/2} g_{_B}$,
with
$h$ being the Planck constant,
$m\sube{}$ the electron mass, $m_{\rm p}$ the proton mass,
$\mu$ the mean molecular weight of the gas
and $g_{_B}$ the Gaunt factor 
(see Rybicki \& Lightman 1979).
The numerical value of ${\mathcal C}$ 
is $3.9\times10^{16}$ in c.g.s. units, assuming that $\mu = 0.5$ and 
$g_{_B}\approx 1$.
The indices $\alpha$ and $\beta$ are constants describing the functional 
form of the second cooling process;
they adopt particular values for radiative accretion shocks
in different physical contexts,
but we retain them as parameters
for the sake of the general formulation,
until results for mCVs are outlined below.
(In general for a cooling function $\Lambda_2\propto\rho^a T^b$,
$\alpha=b-1/2$ and $\beta=3/2-a+b$.)

Expressing the composite cooling function in terms of 
the bremsstrahlung cooling assures us that
the second cooling function
goes to zero as the density becomes infinite.
Using this form assures us that the stability results
depend only on the indices of the cooling function
and are insensitive to
the implementation of the boundary condition at the white dwarf surface,
which becomes unambiguous.
Alternatively,
the cooling could be stopped at a finite density or temperature,
with a softened form of the bremsstrahlung term
and a different choice of lower boundary.
However this approach is mathematically equivalent
to the one we have taken
except that the new boundary conditions
are less obvious,
especially for the perturbed variables.

\section{Perturbation Analysis}
\label{'perturbation'}

The shock position $x\subs{}$
and velocity $v\subs{}$
are subject to a first-order perturbation
(as in Imamura \etal 1996):
\begin{equation}
v\subs{}=v\subs{1} e^{iky+\omega t}
\end{equation}
\begin{equation}
x\subs{}=x\subs{0}+x\subs{1} e^{iky+\omega t}.
\end{equation}
with frequency $\omega$
and a transverse wavenumber $k$ allowing a transverse component
for corrugated oscillation.
Subscripts $0$ and $1$ correspond to the stationary state solution
and perturbed quantities respectively.
In the stationary solution, the shock is at rest:
$v\subs{0}=0$.
The longitudinal perturbation of the shock position and velocity
are related  by $v\subs{1}=x\subs{1}\omega$.
The frequency of the perturbation is $\omega$,
and $k$ is the wavenumber of a transverse component of the 
perturbation in the $y$ direction.
Eigenmodes are labelled by
a transverse wavenumber
and dimensionless eigenfrequency,
which are defined as:
\begin{eqnarray}
\kappa=x\subs{0} k
\\
\delta={{x\subs{0}}\over{\vff}}\omega
\end{eqnarray}
The eigenfrequency is complex,
$\delta=\dR+i\dI$,
where $\dI$ is the oscillatory part
proportional to the frequency of the particular mode.
Stability is indicated by the sign of the growth/decay term $\dR$.
Positive $\dR$ values indicate instability; negative values 
indicate stability.
The physical scales of the system are
$\vff$ the free-fall velocity of the white dwarf,
$\rhoa$ the accretion density,
and $x\subs{0}$ the equilibrium shock height.
The size of the perturbation is indicated by the parameter
$\varepsilon\equiv v\subs{1}/\vff$,
and therefore $x\subs{1}=\varepsilon x\subs{0} /\delta$.

To simplify the calculations,
the hydrodynamic quantities are expressed in terms of dimensionless 
variables,
allowing the scales of the system to be removed from later equations.
These dimensionless variables are expressed in terms of
the stationary solutions and perturbations of
the velocity, density, total pressure and electron pressure,
varying with position in the accretion column
$\xi\equiv x/x\subs{}$.
The white dwarf surface is at $\xi=0$
and the shock is $\xi=1$.
\begin{eqnarray}
\rho(\xi,y,t)&=&\rhoa\cdot
\zeta_0(\xi) \left({1+\varepsilon\lambda_\zeta(\xi) e^{iky+\omega t} }\right)
\\
\vb(\xi,y,t)&=&-\vff\cdot
\left({
\tau_0(\xi) \left({1+\varepsilon\lambda_\tau(\xi) e^{iky+\omega t}}\right),
\varepsilon\tau_0(\xi)\lambda_y(\xi) e^{iky+\omega t}
}\right)
\\
P(\xi,y,t)&=&\rhoa\vff^2\cdot
\pi_0(\xi) \left({1+\varepsilon\lambda_\pi(\xi) e^{iky+\omega t}}\right)
\\
P\sube{}(\xi,y,t)&=&\rhoa\vff^2\cdot
\pi\sube{0}(\xi) \left({1+\varepsilon\lambda\sube{}(\xi) e^{iky+\omega t}}\right)
.
\end{eqnarray}
where the complex functions $\lambda_\zeta$, $\lambda_\tau$, $\lambda_y$,
$\lambda_\pi$ and $\lambda\sube{}$
represent the effect of the perturbation on the downstream profiles of
density, vertical velocity, transverse velocity, total pressure
and electron pressure respectively.

The space and time derivatives of a flow variable $f$,
in a frame following the shock,
are given by:
\begin{eqnarray}
{{Df}\over{Dx}}
&\approx&
{1\over{x\subs{0}}}\cdot
\left({
1-\varepsilon {{e^{iky+\omega t}}\over\delta}
}\right)
\left({
{\partial f}\over{\partial\xi}
}\right)
\\
{{Df}\over{Dy}}
&\approx&
{{\partial f}\over{\partial y}}
-{1\over{x\subs{0}}}\cdot i\varepsilon {{\xi\kappa}\over{\delta}} e^{iky+\omega t}
\left({
{\partial f}\over{\partial\xi}
}\right)
\\
{{Df}\over{Dt}}
&\approx&
{{\partial f}\over{\partial t}}
-{{\vff}\over{x\subs{0}}}\cdot
\varepsilon \xi e^{iky+\omega t}
\left({
{\partial f}\over{\partial\xi}
}\right).
\end{eqnarray}
None of the hydrodynamic variables has an explicit dependence
on position within the accretion column.
These shock-frame derivatives are used to expand
the equations for mass, momentum and energy continuity,
whence are extracted expressions describing the stationary solution
and a set of differential equations for the perturbed variables.

We consider expressions in terms of the dimensionless stationary-solution
velocity ($\tau_0\equiv-v_0/\vff$) to simplify the form of the 
equations.
The stationary solution is obtained by
solving the hydrodynamic equations without the time-dependent terms,
and it is completely described by
two simple algebraic equations for the density and pressure,
\begin{equation}
\zeta_0=1/\tau_0
\label{'eq.unitless.mass'}
\end{equation}
\begin{equation}
\pi_0=1-\tau_0
\label{'eq.unitless.pressure'}
\end{equation}
plus two differential equations:
one for the velocity-position profile
and the other for the electron pressure:
\begin{equation}
{{d\xi}\over{d\tau_0}}
={{\gamma\pi_0-\tau_0}\over{\tL}}
\label{'eq.stationary.velocity'}
\end{equation}
\begin{equation}
{{d\pi\sube{}}\over{d\tau_0}}
={1\over{\tau_0}}
\left[{
\left({
1-{\tG\over\tL}
}\right)
(\gamma\pi_0-\tau_0)-\gamma\pi\sube{}
}\right]
\label{'eq.stationary.pressure'}
\end{equation}
where
\begin{equation}
\tG=(\gamma-1)(\rhoa\vff^3/x\subs{0})^{-1}\Gamma
=(\gamma-1)\psic\psiei
{ {1-\tau_0-2\pi\sube{}} \over { \sqrt{\tau_0^5\pi\sube{}^3} } }
\label{'eq.bigG'}
\end{equation}
and
\begin{equation}
\tL=(\gamma-1)
(\rhoa\vff^3/x\subs{0})^{-1}
\Lambda
=(\gamma-1)\psic\sqrt{{\pi\sube{}}\over{\tau_0^3}}
\left[{
1+\epsilon\subs{}f\left({\tau_0,\pi\sube{}}\right)
}\right]
\label{'eq.bigL'}
\end{equation}
are appropriate dimensionless forms
of the electron-ion exchange and total cooling function respectively.
The parameter $\psiei$ describes the speed of the electron-ion exchange
compared to the cooling;
when $\psiei\LS 1$ the two-temperature treatment is justified;
otherwise the system behaves like the single-temperature shock
described in Saxton \etal \shortcite{saxton98}.
In physical terms the constants $\psic$ and $\psiei$ are:
\begin{equation}
\psic={{x\subs{0}\rhoa}\over\vff}\cdot
{\mathcal C},
\end{equation}
and
\begin{equation}
\psiei=
{1\over{\vff^2}}\cdot
{{(\gamma-1)}\over{\mathcal C}}
4(2\pi)^{1/2}\ln C
\left({{e^4 m\sube{}^{1/2}}\over{m\subi{}^{7/2} }}\right)
\Upsilon(\sigma\subs{}).
\end{equation}
and $\Upsilon(\sigma\subs{})\approx1$ (see Saxton 1999).
The function $f(\tau_0,\pi\sube{})$ describes
the local ratio of the power of the second cooling process
relative to the bremsstrahlung emission.
It takes the form:
\begin{equation}
f(\tau_0,\pi\sube{})={{4^{\alpha+\beta}}\over{3^\alpha}}
\left({
{1+\sigma\subs{}}\over{\sigma\subs{}}
}\right)^{\alpha}
\pi\sube{}^\alpha \tau_0^\beta
\label{'eq.define.f'}
\end{equation}
and $\sigma\subs{}\equiv (P\sube{}/P\subi{})\subs{}$
is the ratio of electron to ion pressures at the shock surface.
Upon substitution of these expressions,
(\ref{'eq.unitless.mass'},
\ref{'eq.unitless.pressure'}
\ref{'eq.bigG'},\ref{'eq.bigL'},\ref{'eq.define.f'})
the equations for the stationary solution velocity
(\ref{'eq.stationary.velocity'})
and pressure profiles
(\ref{'eq.stationary.pressure'})
become:
\begin{equation}
{{d\xi}\over{d\tau_0}}
=
{
{\gamma-(\gamma+1)\tau_0}
\over{(\gamma-1)\psic\left[{1+\epsilon\subs{}f(\tau_0,\pi\sube{})}\right]}}
\sqrt{{\tau_0^3}\over{\pi\sube{} }}
\end{equation}
\begin{equation}
{{d\pi\sube{}}\over{d\tau_0}}=
{1\over{\tau_0}}
\left[{
\left({
1-{ {\psiei}\over{1+\epsilon\subs{}f(\tau_0,\pi\sube{})} }
{{1-\tau_0-2\pi\sube{} }\over{\tau_0\pi\sube{}^2}}
}\right)
\left({\gamma-\left({\gamma+1}\right)\tau_0}\right) -\gamma\pi\sube{}
}\right]
\end{equation}
For the case of bremsstrahlung cooling alone 
($\epsilon\subs{}=0$) there exists an analytic solution
(Aizu 1973).
For the general two-process cooling function it is necessary to
carry out a numerical integration.

The hydrodynamic equations expressed in terms of
the dimensionless perturbed variables give rise to
a set of five coupled complex first-order differential equations,
which can be expressed in matrix form as:
\begin{equation}
{
\left[
\begin{array}{rrrrr}
-\tau_0&-\tau_0&0&0&0\\
0&-\tau_0&0&\tau_0-1&0\\
0&0&-\tau_0&0&0\\
\gamma\tau_0\pi_0&0&0&-\tau_0\pi_0&0\\
\gamma\tau_0\pi\sube{}&0&0&0&-\tau_0\pi\sube{}\\
\end{array}
\right]
}
{
\left[
\begin{array}{c}
\lambda_\zeta'\\
\lambda_\tau'\\
\lambda_y'\\
\lambda_\pi'\\
\lambda_{\rm e}'
\end{array}
\right]
}
=
{
\left[
\begin{array}{c}
F_1\\
F_2\\
F_3\\
F_4\\
F_5
\end{array}
\right]
\label{'eq.2d2t.matrix.1'}
}
\end{equation}
where a prime denotes derivatives in $\xi$
and the functions on the right hand side collect the terms which
lack derivatives of the perturbed variables. In general, 
for a cooling function composed of a sum of several power-law terms,
they are:
\begin{eqnarray}
F_1(\tau_0,\pi\sube{},\xi)
&=&
-\xi(\ln \tau_0)'
-\delta\lambda_\zeta
+i\kappa \tau_0\lambda_y
\\
F_2(\tau_0,\pi\sube{},\xi)
&=&
-(\delta-\tau_0')\lambda_\tau
+\xi(\ln \tau_0)'
+\tau_0'\lambda_\zeta
-\tau_0'(\lambda_\pi-\lambda_\tau)
\\
F_3(\tau_0,\pi\sube{},\xi)
&=&
-(\delta-\tau_0')\lambda_y
+i\kappa(1-\tau_0)\lambda_\pi
+i\kappa \tau_0'\xi/\delta
\\
F_4(\tau_0,\pi\sube{},\xi)
&=&
-\pi_0\delta(\lambda_\pi -\gamma\lambda_\zeta)
-\tL\left[{
-{3\over2}g_2(\tau_0,\pi\sube{})\lambda_\zeta
+{1\over2}g_1(\tau_0,\pi\sube{})\lambda\sube{}
-\lambda_\tau-\lambda_\pi+{1\over\delta}-{\xi\over{\tau_0}}
}\right]
\\
F_5(\tau_0,\pi\sube{},\xi)
&=&
-\pi\sube{}\delta(\lambda\sube{} -\gamma\lambda_\zeta)
-\tL\left[{
-{3\over2}g_2(\tau_0,\pi\sube{})\lambda_\zeta
+{1\over2}g_1(\tau_0,\pi\sube{})\lambda\sube{}
-\lambda_\tau-\lambda_\pi+{1\over\delta}-{\xi\over{\tau_0}}
}\right]
\nonumber \\
& &+\tG \left[{
{5\over2}\lambda_\zeta -{3\over2}\lambda\sube{}
+{{\pi_0\lambda_\pi -2\pi\sube{}\lambda\sube{} }\over{\pi_0-2\pi\sube{}}}
-\lambda_\tau-\lambda\sube{}+{1\over\delta}-{\xi\over{\tau_0}}
}\right]
\end{eqnarray}
where the functions $g_1(\tau_0,\pi\sube{})$ and $g_2(\tau_0,\pi\sube{})$
are defined as:
\begin{equation}
g_1(\tau_0,\pi\sube{}) = 1
+{{2{\epsilon_{\rm s}}\alpha f(\tau_0,\pi\sube{})}
\over{1+\epsilon_{\rm s}f(\tau_0,\pi\sube{})}},
\end{equation}
\begin{equation}
g_2(\tau_0,\pi\sube{}) = 1-{2\over 3} \left[{
{{{\epsilon_{\rm s}}\beta f(\tau_0,\pi\sube{})}
\over{1+\epsilon_{\rm s}f(\tau_0,\pi\sube{})}}
 }\right],
\end{equation}

The perturbed transverse velocity variable $\lambda_y$
has an inconspicuous influence.
If not for its presence in the $F_1$ function,
$\lambda_y$ would be decoupled from the other perturbed variables,
and could be integrated separately after finding the solutions
in the other variables.

The matrix in
(\ref{'eq.2d2t.matrix.1'})
is non-singular everywhere in the post-shock region
except at $\tau_0=0$.
In a practice this singularity doesn't matter
because the numerical integration can be taken to some small value
$\tau_0=\Delta>0$ and the results reach a limit as
$\Delta\rightarrow 0$.
The matrix equation
is inverted using (\ref{'eq.stationary.velocity'})
to yield equations for the perturbed variables.
\begin{equation}
{d\over{d\tau_0}}
{
\left[
\begin{array}{c}
\lambda_\zeta\\
\lambda_\tau\\
\lambda_y\\
\lambda_\pi\\
\lambda\sube{}
\end{array}
\right]
}
=
{1\over\tL}
{
\left[
\begin{array}{ccccc}
1&-1&0&{1/{\tau_0}}&0\\
{-{\gamma\pi_0}/{\tau_0}}&1&0&-{1/{\tau_0}}&0\\
0&0&-{{(\gamma\pi_0-\tau_0)}/{\tau_0}}&0&0\\
\gamma&-\gamma&0&{1/{\pi_0}}&0\\
\gamma&-\gamma&0&{\gamma/{\tau_0}}&-{{(\gamma\pi_0-\tau_0)}/{\tau_0\pi\sube{}}}
\end{array}
\right]
}
{
\left[
\begin{array}{c}
F_1\\
F_2\\
F_3\\
F_4\\
F_5\\
\end{array}
\right]
}
\label{'eq.2d2t.matrix.2'}
\end{equation}
These resulting differential equations are split into real and imaginary 
parts to yield ten first-order differential equations in real eigenfunctions,
${\lambda_\zeta}\subR$,
${\lambda_\zeta}\subI$,
${\lambda_\tau}\subR$,
${\lambda_\tau}\subI$,
${\lambda_y}\subR$,
${\lambda_y}\subI$,
${\lambda_\pi}\subR$,
${\lambda_\pi}\subI$,
${\lambda\sube{}}\subR$,
and ${\lambda\sube{}}\subI$,
plus the known parameters and variables of the static solution,
$\alpha$, $\beta$, $\xi(\tau_0)$ and $\pi\sube{}(\tau_0)$.
This decomposition is straightforward because only the vector parts of
(\ref{'eq.2d2t.matrix.2'}) are complex.
To obtain the equations for real variables,
we substitute the real parts of the $\lambda$-variables in the left side
and real parts of the $F$-functions in the right side.
Similarly, the equations for the eigenfunctions' imaginary parts
are obtained by substituting imaginary parts of
the $\lambda$-variables and $F$-functions.

Optically thick cyclotron emission provides an important cooling process
for accretion onto strongly magnetic white dwarfs.
An optically thick radiative process cannot in general be described by
a simple local energy loss function.
A full treatment requires simultaneous solution to
the time-dependent hydrodynamic and radiative transfer equations.
However for the particular geometry of the field-aligned accretion flow
in mCVs and the physical conditions typical of such systems,
cyclotron radiation is optically thick up to a critical frequency
at which it becomes optically thin.
The total power escaping from the post-shock region locally
in the accretion column approximately takes the form of a power-law 
cooling term $\Lambda_{\rm cyc}\propto\rho^{3/20}T\sube{}^{5/2}$,
which in the terms of (\ref{'eq.lambda.total'})
corresponds to indices $\alpha=2.0$ and $\beta=3.85$
(see Langer \etal 1982,  Wu, Chanmugam \& Shaviv 1994,
Cropper \etal 1999).
This approximation renders the linear analysis tractable.

For the special case of a mCV radiative shock with
bremsstrahlung and cyclotron cooling present,
the parameter $\epsilon\subs{}$ is related to
the magnetic field at the accretion pole.
The expression for $\epsilon\subs{}$ is as described in Wu (1994).
In the single-temperature limit $\sigma\subs{}=1$,
but the value in the two-temperature conditions
depends on the conduction of energy
upstream of the ion shock, carried by the electrons.
Our analysis treats $\sigma\subs{}$ as a parameter.

\begin{table}
\caption{Eigenvalues $\dR$, $\dI$
in the absence of transverse perturbation ($\kappa=0$),
for various conditions of electron/ion pressure ratio at the shock
($\sigma\subs{}$),
strength of electron-ion energy exchange ($\psiei$),
and relative efficiency of cyclotron cooling ($\epsilon\subs{}$).
}
\hspace{20cm}
\begin{center}
$
\begin{array}{rrrr}
\hline
\begin{array}{rr}\sigma\subs{}&\psiei\end{array}&
\epsilon\subs{}=0~~~~
&\epsilon\subs{}=1~~~~
&\epsilon\subs{}=100~~~~
\\
\hline
\begin{array}{rr}0.2&0.1\\ \\ \\ \\ \\ \\\end{array}&
\begin{array}{rr}
-0.001&0.307\\
 0.048&0.902\\
 0.066&1.529\\
 0.092&2.141\\
 0.095&2.769\\
 0.114&3.385\\
\end{array}
&
\begin{array}{rr}
-0.057&0.343\\
-0.040&0.808\\
-0.030&1.363\\
 0.005&1.896\\
-0.001&2.444\\
 0.027&2.995\\
\end{array}
&
\begin{array}{rr}
-0.113&0.335\\
-0.073&0.567\\
-0.067&0.978\\
-0.052&1.387\\
-0.036&1.758\\
-0.046&2.154\\
\end{array}
\\
\\
\begin{array}{rr}0.2&0.5\\ \\ \\ \\ \\ \\\end{array}&
\begin{array}{rr}
-0.007&0.306\\
 0.047&0.894\\
 0.060&1.513\\
 0.085&2.120\\
 0.087&2.739\\
 0.106&3.349\\
\end{array}
&
\begin{array}{rr}
-0.106&0.340\\
-0.054&0.758\\
-0.076&1.272\\
-0.040&1.756\\
-0.054&2.253\\
-0.026&2.770\\
\end{array}
&
\begin{array}{rr}
-0.166&0.247\\
-0.043&0.528\\
-0.111&0.775\\
-0.088&1.104\\
-0.071&1.406\\
-0.086&1.710\\
\end{array}
\\
\\
\begin{array}{rr}0.2&1.0\\ \\ \\ \\ \\ \\\end{array}&
\begin{array}{rr}
-0.009&0.306\\
 0.047&0.891\\
 0.060&1.509\\
 0.085&2.115\\
 0.086&2.733\\
 0.105&3.342\\
\end{array}
&
\begin{array}{rr}
-0.123&0.332\\
-0.056&0.744\\
-0.087&1.243\\
-0.046&1.717\\
-0.067&2.193\\
-0.043&2.701\\
\end{array}
&
\begin{array}{rr}
-0.163&0.211\\
-0.056&0.520\\
-0.101&0.701\\
-0.109&1.006\\
-0.081&1.271\\
-0.094&1.541\\
\end{array}
\\
\\
\begin{array}{rr}0.5&0.1\\ \\ \\ \\ \\ \\\end{array}&
\begin{array}{rr}
 0.003&0.304\\
 0.051&0.898\\
 0.068&1.522\\
 0.093&2.132\\
 0.096&2.756\\
 0.114&3.370\\
\end{array}
&
\begin{array}{rr}
-0.025&0.322\\
 0.006&0.849\\
 0.020&1.435\\
 0.044&2.002\\
 0.044&2.585\\
 0.064&3.161\\
\end{array}
&
\begin{array}{rr}
-0.057&0.309\\
-0.027&0.582\\
-0.014&1.013\\
 0.003&1.416\\
 0.002&1.819\\
 0.008&2.232\\
\end{array}
\\
\\
\begin{array}{rr}0.5&0.5\\ \\ \\ \\ \\ \\\end{array}&
\begin{array}{rr}
-0.006&0.306\\
 0.048&0.892\\
 0.061&1.511\\
 0.086&2.117\\
 0.088&2.735\\
 0.107&3.345\\
\end{array}
&
\begin{array}{rr}
-0.065&0.332\\
-0.019&0.823\\
-0.016&1.389\\
 0.011&1.930\\
 0.006&2.490\\
 0.031&3.047\\
\end{array}
&
\begin{array}{rr}
-0.135&0.264\\
-0.023&0.551\\
-0.081&0.868\\
-0.049&1.218\\
-0.051&1.547\\
-0.059&1.906\\
\end{array}
\\
\\
\begin{array}{rr}0.5&1.0\\ \\ \\ \\ \\ \\\end{array}&
\begin{array}{rr}
-0.008&0.305\\
 0.048&0.891\\
 0.061&1.508\\
 0.085&2.113\\
 0.087&2.730\\
 0.106&3.339
\end{array}
&
\begin{array}{rr}
-0.078&0.332\\
-0.023&0.814\\
-0.023&1.376\\
 0.006&1.910\\
-0.002&2.463\\
 0.023&3.015
\end{array}
&
\begin{array}{rr}
-0.148&0.239\\
-0.036&0.549\\
-0.092&0.816\\
-0.073&1.159\\
-0.064&1.448\\
-0.083&1.781
\end{array}
\\
\\
\begin{array}{rr}1.0&0.1\\ \\ \\ \\ \\ \\\end{array}&
\begin{array}{rr}
\ 0.010&0.299\\
 0.056&0.892\\
 0.071&1.511\\
 0.093&2.118\\
 0.095&2.735\\
 0.112&3.345
\end{array}
&
\begin{array}{rr}
-0.001&0.303\\
 0.036&0.856\\
 0.047&1.447\\
 0.064&2.025\\
 0.064&2.610\\
 0.078&3.192
\end{array}
&
\begin{array}{rr}
 0.002&0.277\\
 0.027&0.564\\
 0.048&0.997\\
 0.052&1.378\\
 0.043&1.798\\
 0.060&2.199
\end{array}
\\
\\
\begin{array}{rr}1.0&0.5\\ \\ \\ \\ \\ \\\end{array}&
\begin{array}{rr}
-0.005&0.304\\
 0.049&0.890\\
 0.063&1.507\\
 0.087&2.112\\
 0.089&2.729\\
 0.107&3.337
\end{array}
&
\begin{array}{rr}
-0.039&0.319\\
 0.009&0.846\\
 0.019&1.430\\
 0.042&1.995\\
 0.040&2.576\\
 0.061&3.150
\end{array}
&
\begin{array}{rr}
-0.096&0.262\\
 0.007&0.554\\
-0.032&0.909\\
-0.000&1.257\\
-0.022&1.610\\
-0.014&1.992
\end{array}
\\
\\
\begin{array}{rr}1.0&1.0\\ \\ \\ \\ \\ \\\end{array}&
\begin{array}{rr}
-0.007&0.305\\
 0.049&0.890\\
 0.062&1.506\\
 0.086&2.110\\
 0.088&2.727\\
 0.106&3.335
\end{array}
&
\begin{array}{rr}
-0.050&0.322\\
 0.004&0.843\\
 0.012&1.425\\
 0.038&1.988\\
 0.035&2.566\\
 0.056&3.139\\
\end{array}
&
\begin{array}{rr}
-0.122&0.247\\
-0.015&0.559\\
-0.059&0.876\\
-0.025&1.230\\
-0.038&1.545\\
-0.046&1.915\\
\end{array}
\\
\hline
\end{array}
$
\end{center}
\label{'table.bremcyclotron'}
\end{table}

\section{Results}
\label{'results'}

\subsection{Eigen-modes}

The ten real differential equations are integrated numerically
using a Runge-Kutta method.
In terms of the dimensionless perturbed variables,
the boundary conditions at the shock are $\tau_0\!=1/4$,
$\xi=1$, $\pi\sube{}={\frac34}(1+1/\sigma\subs{})^{-1}$,
$\lambda_\zeta=0$, $\lambda_\tau=-3$, $\lambda_y=3i\kappa/\delta$
and $\lambda_\pi=\lambda\sube{}=2$.
At the dwarf surface ($\tau_0\!=0$, $\xi=0$, $\pi\sube{}=1/2$)
there are no specific conditions on
the values of the perturbed density and pressures,
but the stationary wall condition $\lambda_\tau=0$
applies (see Chevalier \& Imamura 1982; Imamura \etal 1996;
Saxton \etal 1997). Integration proceeds 
between $\tau_0\!=0$ and $\tau_0\!=\!1/4$ for trial values of $\dR$
and $\dI$. Values of the $\delta$'s are sought which yield 
$\lambda_\tau=0$ when the differential equations are integrated to 
$\tau_0=0$ (using the search method described in the appendices of 
Saxton \etal 1998)
Each of these eigenvalues corresponds to an oscillatory mode of the shock.
The modes form an indefinite sequence consisting 
of a fundamental mode and a succession of overtones.

The survey of the complex frequency eigenplane is conducted
in a region of $\delta$ appropriate for the lowest modes.
The range of frequency $\dI$ explored is from 0 to 3.5;
the growth/decay term $\dR$ is taken from $0.4$ down to $-0.4$ or lower 
if necessary in cases of higher $\kappa$.
The perturbed velocity eigenfunction is evaluated at the white dwarf surface
for each sampled of $\delta$.
Values of $-\log|\lambda_\tau|_{\rm wd}$ are plotted and
the maxima of this quantity indicate the $\delta$ eigenvalues.

This procedure is performed for single-temperature
($\sigma\subs{}=1$, large $\psiei$)
and two-temperature
($\sigma\subs{}\le 1$, $\psiei$ finite) cases.
Eigenfrequencies of the six lowest modes are calculated
under the conditions $\sigma\subs{}=0.2, 0.5, 1$ and $\psiei=0.1, 0.5, 1$
to test the effect of the unequal electron and ion pressures at the shock,
and the rate of electron-ion energy exchange in the accretion column.
For each case we investigate the modes' frequencies and stability properties
under conditions with bremsstrahlung cooling alone ($\epsilon\subs{}=0$),
cyclotron cooling comparable to the bremsstrahlung ($\epsilon\subs{}=1$),
and cyclotron cooling dominant ($\epsilon\subs{}=100$),
as evaluated at the shock surface.
These eigenvalues are tabulated in Table~\ref{'table.bremcyclotron'}.

Applying the special restricted choice of the $\epsilon\subs{}=0$
exactly recovers earlier results of Imamura \etal (1996)
for a two-temperature shock with only a single bremsstrahlung cooling process.
Taking the limits of high $\psiei$ and $\sigma\subs{}=1$
reduces our equations to reproduce the single-temperature case
with multiple cooling, as in Saxton \etal (1997).
Taking this limit and $\epsilon\subs{}=0$ together
reproduces the single-temperature bremsstrahlung-cooling only results
of Chevalier \& Imamura (1982).

The frequencies and stability behaviour of the fundamental and first overtone
in the presence of the two-temperature conditions and composite cooling
was also investigated for non-longitudinal perturbations.
This corrugation of the shock is represented by non-zero values of 
the dimensionless transverse wavenumber $\kappa$.

\begin{table}
\caption{
Mode frequency spacing $\dIO$ and offset $\dC$
for a linear fit $\dI\approx\dIO(n-1/2)+\dC$,
under different two-temperature conditions ($\sigma\subs{}$,$\psiei$)
and efficiency of the second cooling process ($\epsilon\subs{}$).
}
\hspace{20cm}
\begin{center}
$
\begin{array}{ccc}
\begin{array}{rrrrr}
\hline
\sigma\subs{}&\psiei&\epsilon\subs{}&\dIO&\dC\\
\hline
0.2&0.1&0&	0.617&-0.013\\
0.2&0.1&1&	0.534& 0.039\\
0.2&0.1&100&	0.374& 0.076\\
\\
0.2&0.5&0&	0.610&-0.010\\
0.2&0.5&1&	0.489& 0.057\\
0.2&0.5&100&	0.294& 0.080\\
\\
0.2&1.0&0&      0.609&-0.011\\
0.2&1.0&1&      0.476& 0.060\\
0.2&1.0&100&    0.263& 0.086\\
\hline
\end{array}
&
\begin{array}{rrrrr}
\hline
\sigma\subs{}&\psiei&\epsilon\subs{}&\dIO&\dC\\
\hline
0.5&0.1&0&	0.615&-0.014\\
0.5&0.1&1&	0.571& 0.014\\
0.5&0.1&100&	0.392& 0.051\\
\\
0.5&0.5&0&	0.609&-0.011\\
0.5&0.5&1&	0.547& 0.030\\
0.5&0.5&100&	0.330& 0.069\\
\\
0.5&1.0&0&	0.608&-0.010\\
0.5&1.0&1&	0.540&0.032\\
0.5&1.0&100&	0.307&0.077\\
\hline
\end{array}
&
\begin{array}{rrrrr}
\hline
\sigma\subs{}&\psiei&\epsilon\subs{}&\dIO&\dC\\
\hline
1.0&0.1&0&	0.610&-0.015\\
1.0&0.1&1&	0.580& 0.000\\
1.0&0.1&100&	0.391& 0.028\\
\\
1.0&0.5&0&	0.608&-0.011\\
1.0&0.5&1&	0.569& 0.013\\
1.0&0.5&100&	0.348& 0.055\\
\\
1.0&1.0&0&	0.608&-0.011\\
1.0&1.0&1&	0.566& 0.015\\
1.0&1.0&100&	0.333& 0.063\\
\hline
\end{array}
\end{array}
$
\end{center}
\label{'table.frequency1'}
\end{table}

\begin{table}
\caption{
Parameters of the frequency spacing for a quadratic fit,
$\dI\approx\dQ n^2+\dIO(n-1/2)+\dC$,
under different two-temperature conditions ($\sigma\subs{}$,$\psiei$)
and efficiency of the second cooling process ($\epsilon\subs{}$).
}
\hspace{20cm}
\begin{center}
$
\begin{array}{ccc}
\begin{array}{rrrrrr}
\hline
\sigma\subs{}&\psiei&\epsilon\subs{}&\dIO&\dC&\dQ\\
\hline
0.2&0.1&0&	0.604&-0.002&0.002\\
0.2&0.1&1&	0.484& 0.080&0.007\\
0.2&0.1&100&	0.291& 0.144&0.012\\
\\
0.2&0.5&0&	0.596& 0.001&0.002\\
0.2&0.5&1&	0.436& 0.102&0.007\\
0.2&0.5&100&	0.252& 0.115&0.006\\
\\
0.2&1.0&0&      0.594& 0.002&0.002\\
0.2&1.0&1&      0.428& 0.100&0.007\\
0.2&1.0&100&    0.245& 0.100&0.003\\
\hline
\end{array}
&
\begin{array}{rrrrrr}
\hline
\sigma\subs{}&\psiei&\epsilon\subs{}&\dIO&\dC&\dQ\\
\hline
0.5&0.1&0&	0.602&-0.033&0.002\\
0.5&0.1&1&	0.541& 0.038&0.004\\
0.5&0.1&100&	0.319& 0.113&0.011\\
\\
0.5&0.5&0&	0.595&0.001&0.002\\
0.5&0.5&1&	0.508&0.062&0.005\\
0.5&0.5&100&	0.279&0.112&0.007\\
\\
0.5&1.0&0&	0.594&0.001&0.002\\
0.5&1.0&1&	0.501&0.065&0.006\\
0.5&1.0&100&	0.282&0.098&0.004\\
\hline
\end{array}
&
\begin{array}{rrrrrr}
\hline
\sigma\subs{}&\psiei&\epsilon\subs{}&\dIO&\dC&\dQ\\
\hline
1.0&0.1&0&	0.601& 0.007&0.001\\
1.0&0.1&1&	0.564& 0.013&0.002\\
1.0&0.1&100&	0.327& 0.082&0.009\\
\\
1.0&0.5&0&	0.595& 0.000&0.002\\
1.0&0.5&1&	0.541& 0.036&0.004\\
1.0&0.5&100&	0.292& 0.101&0.008\\
\\
1.0&1.0&0&	0.593& 0.002&0.002\\
1.0&1.0&1&	0.536& 0.041&0.004\\
1.0&1.0&100&	0.298& 0.093&0.005\\
\hline
\end{array}
\end{array}
$
\end{center}
\label{'table.frequency2'}
\end{table}

\subsection{Frequency quantization}

Under the simple conditions,
for either bremsstrahlung-dominated cooling or a single-temperature shock,
the oscillatory part of the eigenfrequency
follows a quantized sequence like the modes of a pipe open at one end.
In a linear fit to this sequence, the harmonics $n$ follow
$\dI\approx\dIO (n-1/2)+\dC$,
with frequency spacing $\dIO$ and a small constant correction offset $\dC$.
The two-temperature parameters $\sigma\subs{}$ and $\psiei$
and the secondary process' cooling efficiency $\epsilon\subs{}$
determine the values of these constants.
Parameter fits were performed for the first six harmonics
under different conditions,
and the results for the constant and linear coefficients
are shown in Table~\ref{'table.frequency1'}.

For a purely bremsstrahlung-dominated shock the mode spacing $\dIO\approx0.6$
regardless of the two-temperature conditions.
As $\epsilon\subs{}$ is increased
under given $\sigma\subs{}$ and $\psiei$ conditions,
$\dIO$ becomes steadily smaller.
For very efficient cyclotron cooling ($\epsilon\subs{}=100$)
the spacing is reduced to $\dIO\approx0.3$.
The decline of $\dIO$ with increasing $\epsilon\subs{}$
is fastest for cases with high $\psiei$;
and increasing $\psiei$ for given $(\sigma\subs{},\epsilon\subs{})$
(approaching single-temperature conditions)
makes $\dIO$ smaller.
In other words, when the electron-ion exchange is weak,
the effect of $\epsilon\subs{}$ on the mode frequencies
is reduced.

The frequency offset constant $\dC$ increases steadily
with increasing $\epsilon\subs{}$
for all observed values of the two-temperature and cooling parameters
$(\sigma\subs{},\psiei)$.
For an increasingly cyclotron-cooling dominated flow
in the presence of strong two-temperature effects,
$\dC\rightarrow{\frac12}\dIO$,
which means that the nature of the modes becomes more analogous to
a pipe with two open ends.
The increased importance of the two temperature effects
in the large $\epsilon\subs{}$ regime
means that the fixed wall boundary condition loses importance
and the global stability of the accretion column
depends more on the disparity of electron and ion
temperatures within the stream than on the radiative cooling processes
(cf. Imamura \etal 1996).
The frequency
and stability properties in this case are determined
by the exchange process rather than the radiative cooling.

Better fits are obtained by introducing another parameter describing
a quadratic dependence on harmonic number,
$\dI\approx\dQ n^2+\dIO (n-1/2)+\dC$.
This quadratic correction is very small in all examined cases,
but its presence affects the values of $\dIO$ and $\dC$
such that $\dC\rightarrow{\frac12}\dIO$ sooner and more readily
in the high-$\epsilon\subs{}$ cases,
as shown in Table~\ref{'table.frequency2'}.
With this fit $\dC\approx{\frac13}\dIO$ by the point of $\epsilon\subs{}=100$.
Without the $\dQ$ term, the sequence of modes approaches
the behaviour of a doubly-open pipe much more slowly in $\epsilon\subs{}$.

\subsection{Stability properties}

The most unstable systems investigated are those with
bremsstrahlung cooling only ($\epsilon\subs{}=0$).
Varying the values of $\sigma\subs{}$ and $\psiei$ has little effect
in the small-$\epsilon\subs{}$ extreme,
because most of the cooling is due to bremsstrahlung radiation,
which occurs in the dense region
near the white dwarf surface,
where the electron and ions have nearly reached equilibrium
regardless of their initial two-temperature conditions close to the shock.
For most realistic parameters with low or modest $\epsilon\subs{}$,
the fundamental mode is stable.
The fundamental is unstable only in cases of extremely low $\psiei$,
modest or high $\sigma\subs{}$,
and low $\epsilon\subs{}$.
Otherwise the second harmonic
is the lowest potentially unstable mode.

Except when cyclotron cooling and two-temperature effects are both very strong
(when $\psiei$ is very small and $\epsilon\subs{}$ is large),
for given conditions of $(\sigma\subs{},\psiei)$,
increasing $\epsilon\subs{}$ stabilizes every mode in a monotonic manner.
Under these usual conditions
a mode which is unstable for some value of $\epsilon\subs{}$
is also unstable for all cases of lower $\epsilon\subs{}$;
and furthermore an increase in $\epsilon\subs{}$
does not causes a decrease in a mode's stability.

For a given change of $\epsilon\subs{}$
the stability term $\dR$ of each mode progresses
in a way that is individual to the mode
and apparently independent of the stability behaviour of
the other modes of the sequence.
Some modes experience more rapid reduction in $\dR$
for a given increase in $\epsilon\subs{}$,
therefore they tend to stabilize earlier or quicker than other modes,
if all else is equal.

The transition of a particular mode from the unstable ($\dR>0$)
to stable ($\dR<0$) regions of the complex frequency eigenplane
occurs at some threshold value of $\epsilon\subs{}$
that is individual to the mode and dependent on
the $(\sigma\subs{},\psiei)$ conditions.
The lowest modes usually stabilize at lower $\epsilon\subs{}$
because they tend to be in or closer to the $\dR<0$ region
even in the bremsstrahlung-cooling only case.
However there often are exceptions,
because of the the modes' rates of stabilization,
$\partial\dR/\partial\epsilon\subs{}$,
depend on $\epsilon\subs{}$ and are individual to each $n$.
There are many cases of $(\sigma\subs{},\psiei,\epsilon\subs{})$
where a mode $n$ is unstable but a higher mode $n+1$ is stable.
Therefore for some cases of $(\sigma\subs{},\psiei)$
there are modes which can never be the lowest unstable mode
for any value of $\epsilon\subs{}$.

It is instructive to examine collectively
the stability values $\dR$ of the sequence of modes
at some set of parameters $(\sigma\subs{},\psiei,\epsilon\subs{})$.
In the cases of lower $\epsilon\subs{}$ there is a general pattern
in which each successive harmonic is less stable than the lower 
neighbouring mode; $\dR$ increases with harmonic number $n$.
The increases of $\dR$ between consecutive modes
is smaller in the ranges of higher $n$.
As $\epsilon\subs{}$ increases this trend weakens,
because some modes stabilize fast enough to overtake
neighbouring lower modes.
When $\epsilon\subs{}$ is very high,
the modal stability eigenvalues, $\dR$,
are scattered in a negative range of values,
except for the lowest modes
which tend to deviate further towards greater or lesser $\dR$.
The introduction and strengthening of two-temperature effects enhances 
the departures from the general pattern.

\begin{figure}
\begin{center}
\epsfxsize=15.0cm
\epsfbox{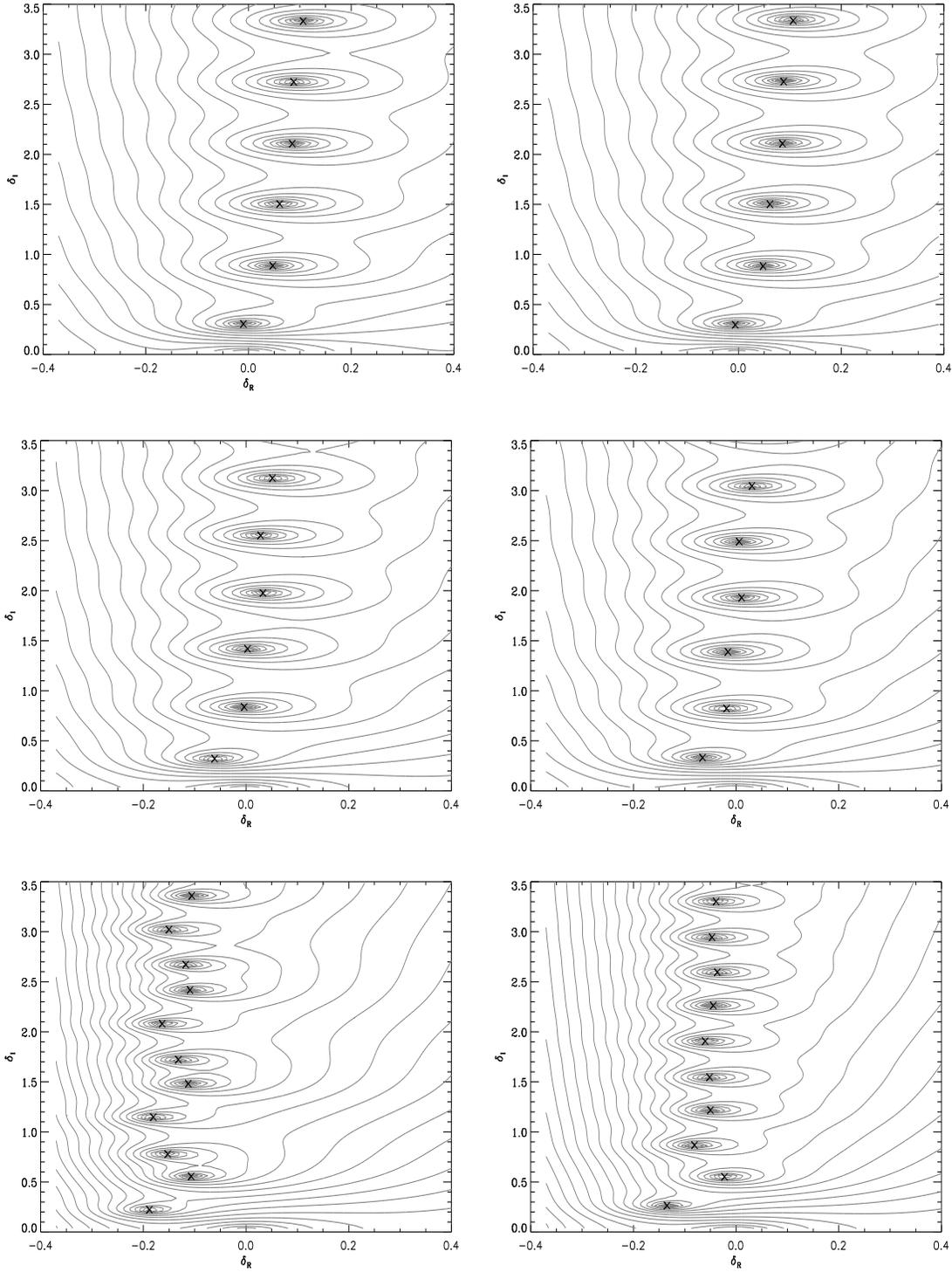}
\end{center}
\caption{
Complex $\delta=\delta\subR+i\delta\subI$ eigenplanes,
with contours of the logarithm of the perturbed velocity eigenfunction
evaluated at the white dwarf surface.
The modes are the points where the perturbed velocity goes to zero,
marked with crosses.
The left column represents the one-temperature cases;
the right column shows two-temperature cases
with $\psiei=0.5$ and $\sigma\subs{}=0.5$.
The top row shows bremsstrahlung-cooling only shocks,
with $\epsilon\subs{}=0$;
in the middle row bremsstrahlung and cyclotron cooling are comparable
$\epsilon\subs{}=1$;
and in the bottom row cyclotron cooling dominates
$\epsilon\subs{}=100$.
}
\label{'figure.2t.eigenplane'}
\end{figure}

The eigenmodes for various $\epsilon\subs{}$
under the one- and two-temperature conditions
are compared in Figure~\ref{'figure.2t.eigenplane'}.
Decreasing $\psiei$ from indefinitely high values
(single-temperature flow) to very low values
(the two-temperature conditions strong)
has little effect on the stability
of the modes if bremsstrahlung cooling is the dominant process
($\epsilon\subs{}$ is small).
In intermediate ranges of $\epsilon\subs{}$,
with cyclotron and bremsstrahlung cooling comparable in efficiency,
strengthening the two-temperature conditions
tends to stabilize most modes.
When $\epsilon\subs{}$ is very high,
the influence of two-temperature effects on the stability of individual 
modes is pronounced.
Introduction of two-temperature effect in high $\epsilon\subs{}$
tends to destabilize most modes,
and the pattern of variations of $\dR$ from one mode to the next
becomes very different from the one-temperature case.
For some $(\sigma\subs{},\psiei)$, two-temperature effects cause
one or more of the lower modes to remain unstable
even for very high $\epsilon\subs{}$,
or even to destabilize a mode as $\epsilon\subs{}$ increases past
some threshold.

\begin{figure}
\begin{center}
\epsfxsize=15cm
\epsfbox{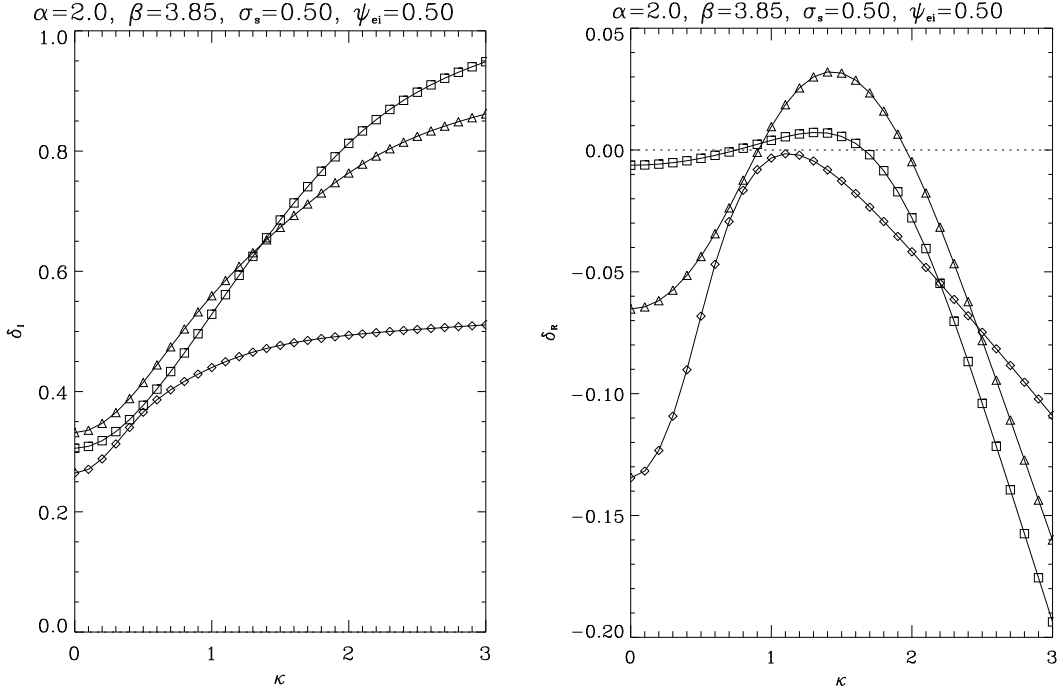}
\end{center}
\caption{Oscillatory part and stability term
of the eigenfrequency for the $n=1$ mode,
with varying transverse wavenumber $\kappa$,
in the presence of two-temperature effects.
Squares, triangles and diamonds
represent the bremsstrahlung plus cyclotron cooling 
cases $\epsilon\subs{}=0,1,100$ respectively.
}
\label{'figure.2t.kappa.n1x'}
\end{figure}

\begin{figure}
\begin{center}
\epsfxsize=15cm
\epsfbox{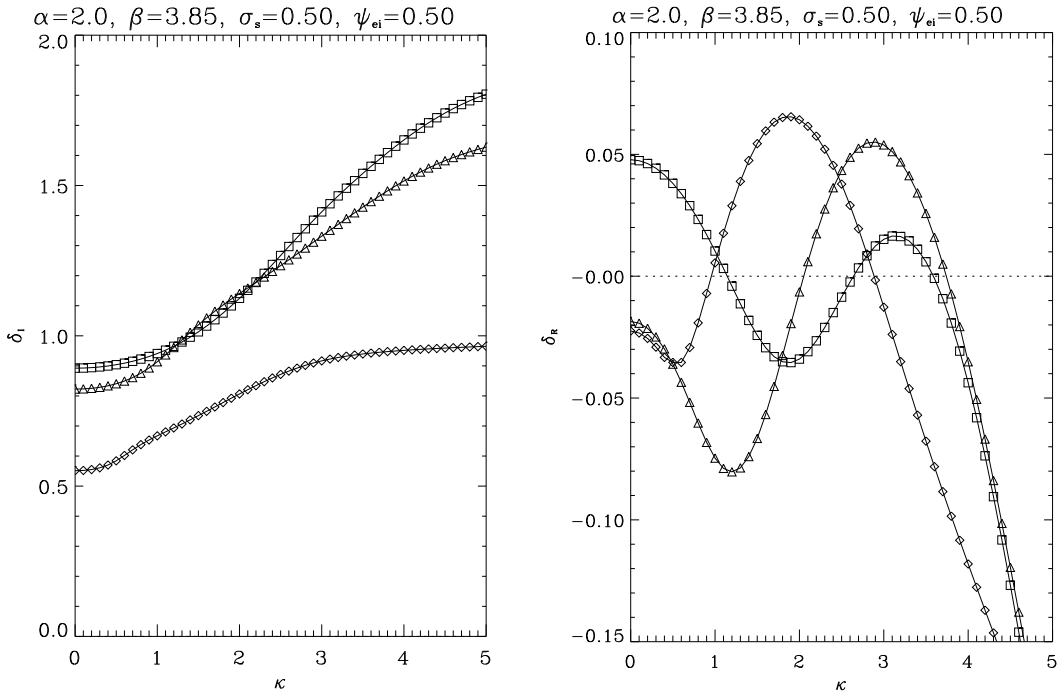}
\end{center}
\caption{Same as Figure~\ref{'figure.2t.kappa.n1x'}
but for the $n=2$ mode.
}
\label{'figure.2t.kappa.n2x'}
\end{figure}

\begin{figure}
\begin{center}
\epsfxsize=15cm
\epsfbox{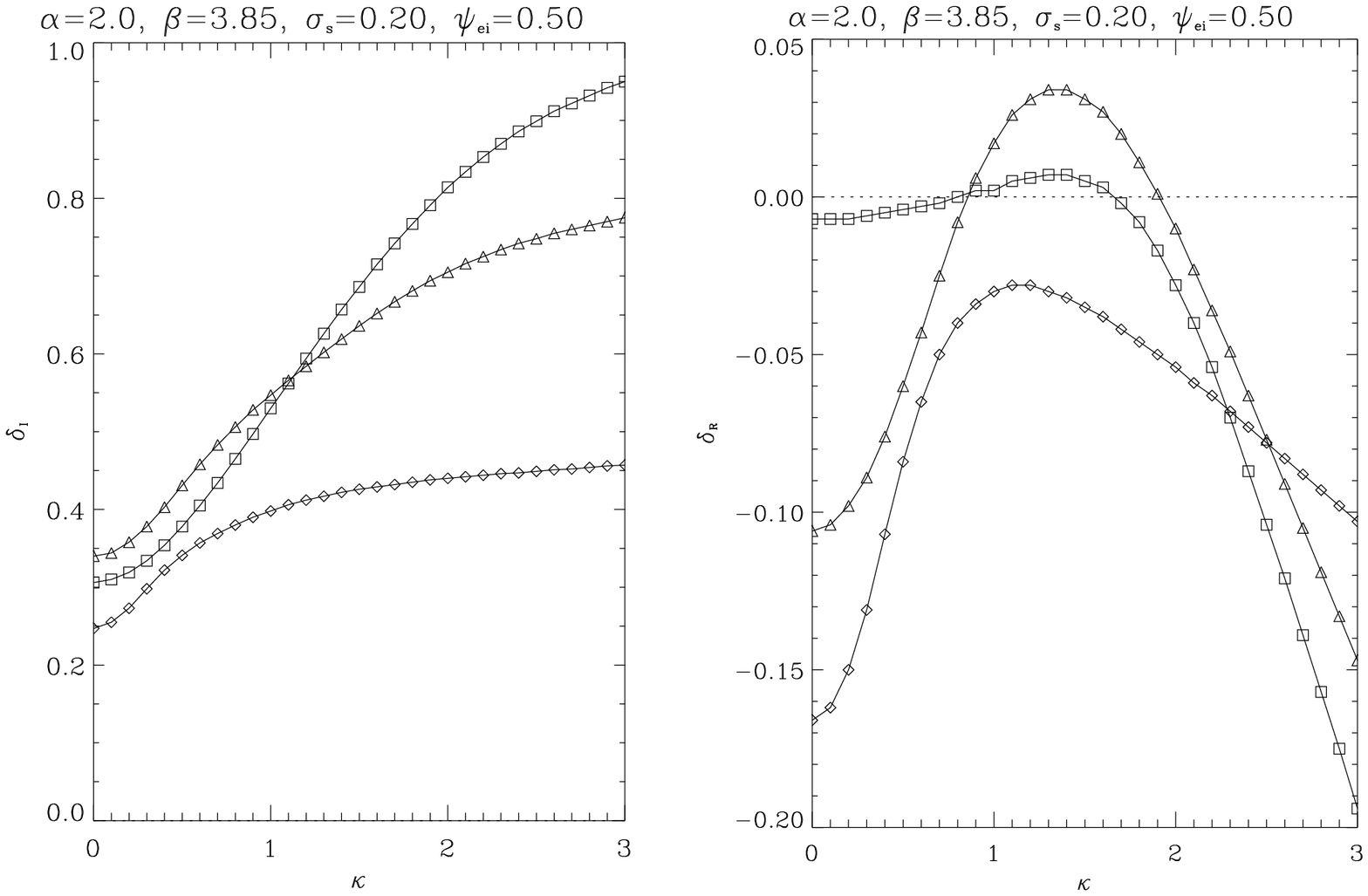}
\end{center}
\caption{Same as Figure~\ref{'figure.2t.kappa.n1x'}
but for the $n=1$ mode,
$\sigma\subs{}=0.2$, $\psiei=0.5$.
}
\label{'figure.2t.kappa.n1y'}
\end{figure}

\begin{figure}
\begin{center}
\epsfxsize=15cm
\epsfbox{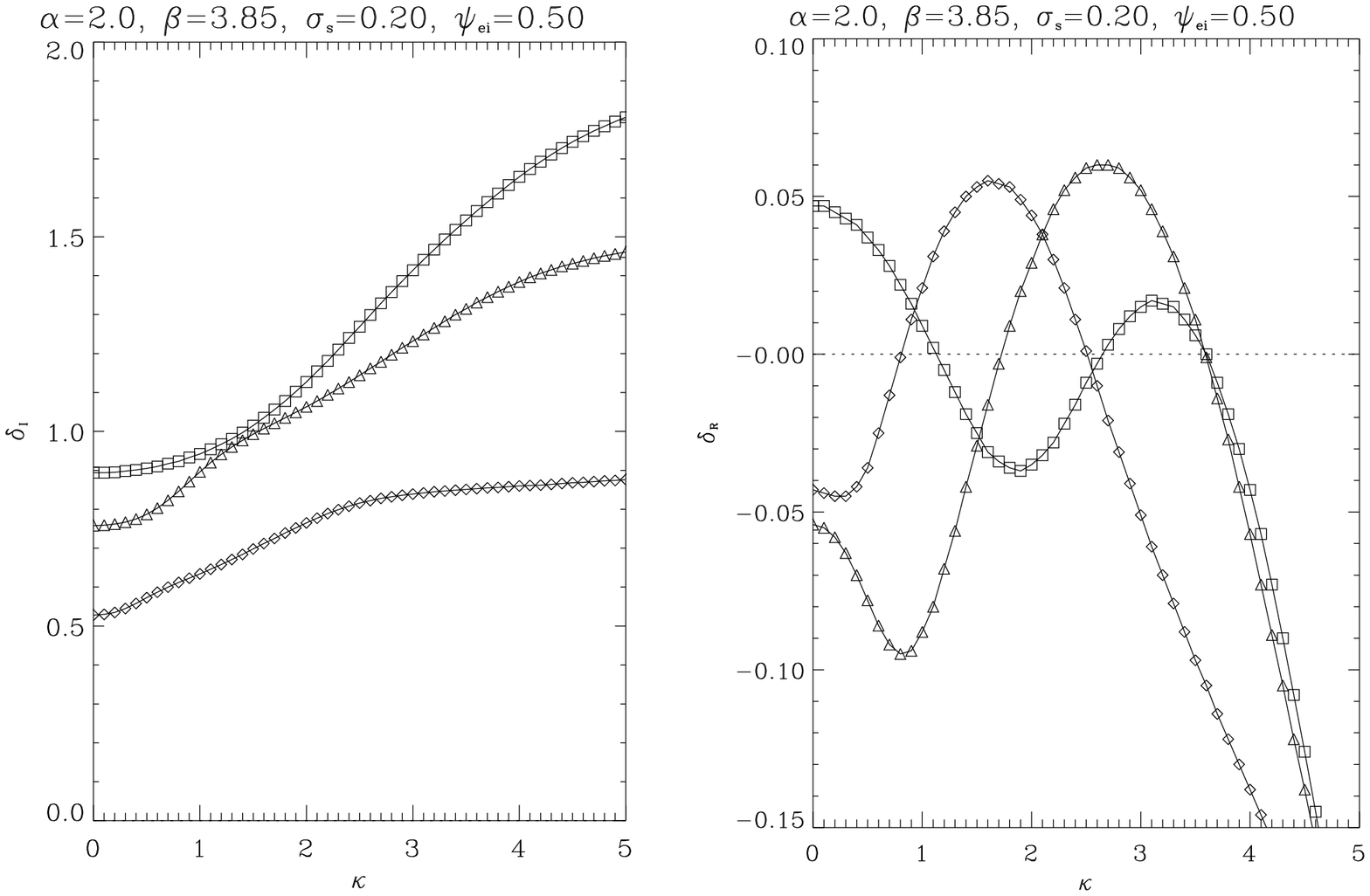}
\end{center}
\caption{Same as Figure~\ref{'figure.2t.kappa.n1x'}
but for the $n=2$ mode,
$\sigma\subs{}=0.2$, $\psiei=0.5$.
}
\label{'figure.2t.kappa.n2y'}
\end{figure}

\begin{figure}
\begin{center}
\epsfxsize=15cm
\epsfbox{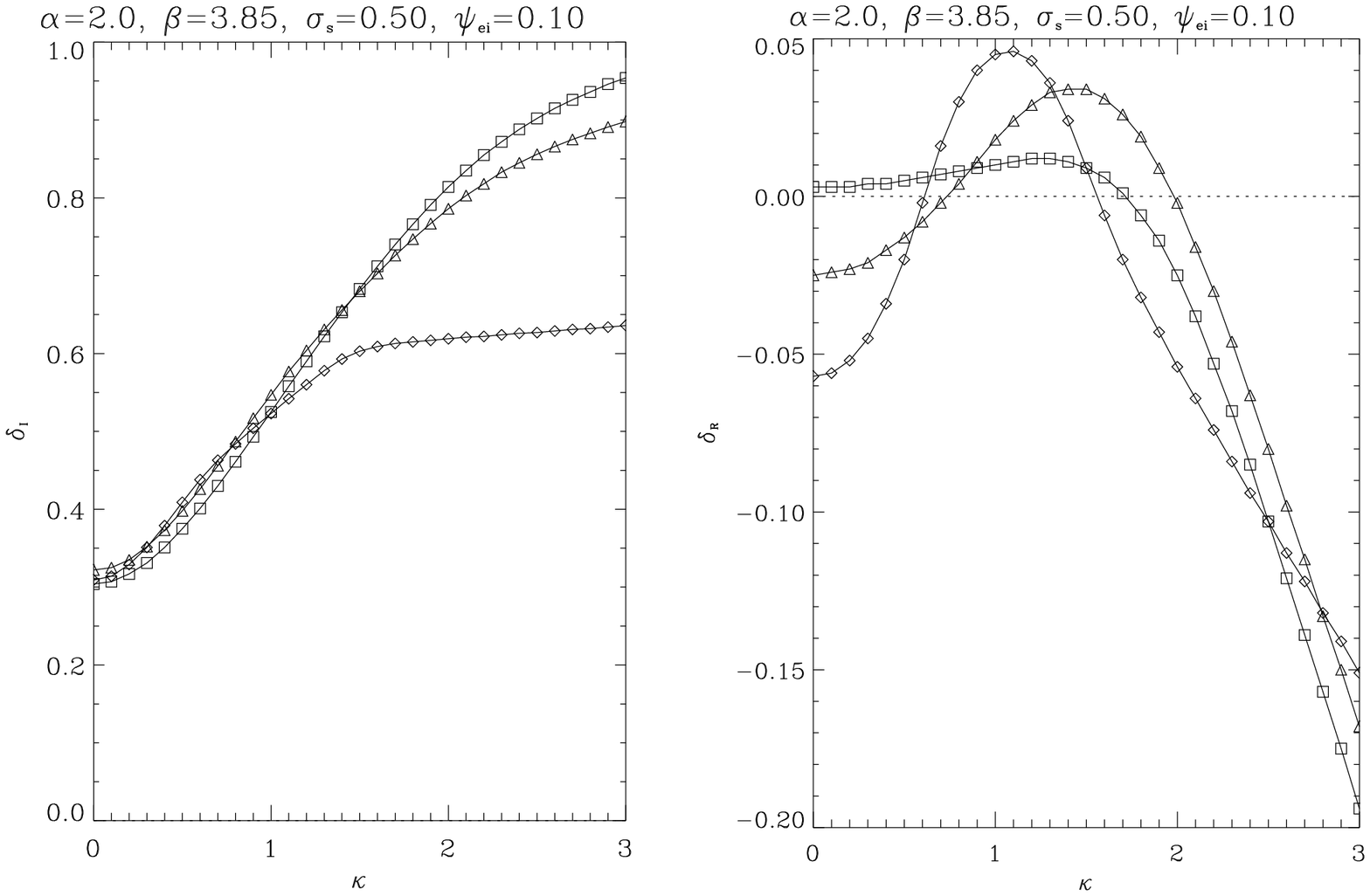}
\end{center}
\caption{Same as Figure~\ref{'figure.2t.kappa.n1x'}
but for the $n=1$ mode,
$\sigma\subs{}=0.5$, $\psiei=0.1$.
}
\label{'figure.2t.kappa.n1z'}
\end{figure}

\begin{figure}
\begin{center}
\epsfxsize=15cm
\epsfbox{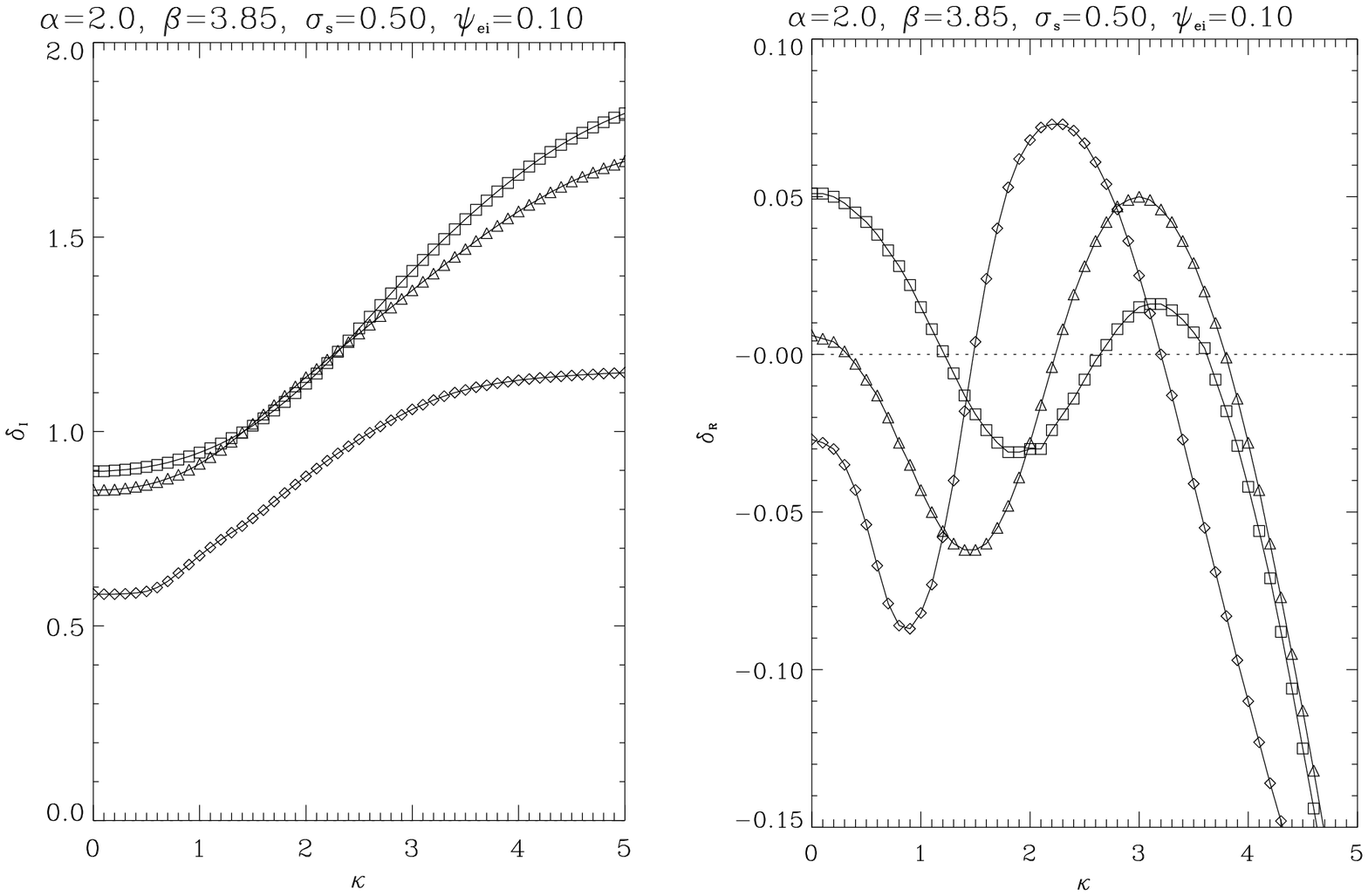}
\end{center}
\caption{Same as Figure~\ref{'figure.2t.kappa.n1x'}
but for the $n=2$ mode,
$\sigma\subs{}=0.5$, $\psiei=0.1$.
}
\label{'figure.2t.kappa.n2z'}
\end{figure}

\subsection{Transverse perturbation}

The stability properties and frequency behaviour of
the oscillatory modes were investigated in the presence of
transverse perturbations, under the two-temperature conditions
and with explicit competition between bremsstrahlung cooling 
and the second cooling (cyclotron cooling).
The eigenvalues of the fundamental and first overtone
were traced through a range of $\kappa$
for particular choices of $(\sigma\subs{},\psiei,\epsilon\subs{})$.

The special case $\kappa=0$ is the absence of transverse perturbations
in which the oscillatory behaviour is the same as
that described in subsections above.
In the limit of large $\kappa$ all of the modes become strongly stable,
and the frequencies approach a sequence like a doubly-open pipe,
$\dI\propto\dIO n$,
Between these extremes, the effect of transverse perturbations
on the frequency and stability of the accretion column
is specific to each mode.
For a given mode with harmonic number $n$,
the interesting behaviour takes place up to a threshold $\kappa\approx 2n+1$.
As $\kappa$ increases, there is a growing region of the eigenplane
within which the modes $n \LS (\kappa-1)/2$
show the simple behaviour like the large-$\kappa$ extreme.
The modes outside the region
experience individual and non-simple disturbances as $\kappa$ is varied,
reverting to the standard $\kappa=0$ behaviour in the large-$n$ modes.

For each mode, the instability term $\dR$
experiences a number of maxima as $\kappa$ varies.
For a mode with harmonic number $n$, there are $n$ of these maxima.
Increasing $\epsilon\subs{}$ affects the values of these maxima
in $(\kappa,\dR)$.
For low $\epsilon\subs{}$ the maxima are most distinct,
but one or more maxima may 
become narrow and shallow almost to the point of vanishing
as $\epsilon\subs{}$ becomes very large. (See Figure 5, for example.)
In the fundamental mode, the single maximum is at about $\kappa\approx3/2$
for low $\epsilon\subs{}$
(see the $\epsilon\subs{}=0$ restricted case in Imamura \etal 1996)
and gradually moves towards lower $\kappa$ as $\epsilon\subs{}$ increases.
In the first overtone, there is a maximum at $\kappa=0$
and another one in the vicinity of $\kappa\sim1-4$.
As $\epsilon\subs{}$ increases, the former maximum lowers in $\dR$ and narrows,
whilst the second maximum moves to lower $\kappa$.
When a maximum of instability (other than $\kappa=0$)
migrates to smaller $\kappa$ with increasing $\epsilon\subs{}$,
the tendency is for the peak $\dR$ to increase
until some critical value of $\epsilon\subs{}$ is attained,
after which the peak $\dR$ decreases slowly with increasing $\epsilon\subs{}$.
These thresholds depend upon $n$, $\psiei$ and $\sigma\subs{}$.
The threshold is less than $\epsilon\subs{}=100$ when $\sigma\subs{}$ is small.
In instances of very high $\epsilon\subs{}$
a mode may be stable for all $\kappa$,
except when $\sigma\subs{}$ is large or $\psiei$ is small.
In all cases studied, enormous values of $\epsilon\subs{} \GS 100$
are required to stabilize modes other than the fundamental.
For example, the fundamental is stabilized completely by $\epsilon\subs{}=100$
in Figures~\ref{'figure.2t.kappa.n1x'},~\ref{'figure.2t.kappa.n1y'},
but not in Figure~\ref{'figure.2t.kappa.n1z'}, where $\psiei$ is too small.

For all modes under all conditions,
the oscillatory frequency $\dI$ increases monotonically with $\kappa$.
In the presence of transverse oscillations,
consecutive modes are no longer unique in $\dI$:
it is possible for modes $n$ and $n+1$
to have the same $\dI$ but different transverse component $\kappa$.
However the modes do not become degenerate;
there is no case in which different modes have the same $\kappa$ and $\dI$.
For each mode the rise of $\dI$ with $\kappa$ is smooth
and is a function of $\epsilon\subs{}$.
Typical values of the gradient $\partial\dI/\partial\epsilon\subs{}$,
for given $\kappa$,
are greater for higher $\epsilon\subs{}$.
For sufficiently high $\kappa$ and $\epsilon\subs{}$,
$\dI$ reaches a plateau for each mode.
(See the left side of Figures~\ref{'figure.2t.kappa.n1x'}-\ref{'figure.2t.kappa.n2z'}).

\subsection{Summary}

We have found:
\begin{enumerate}
\item
When $\epsilon\subs{}$ is large,
two-temperature effects are very important in determining
the frequency and stability properties of the accretion flow;
when $\epsilon\subs{}$ is small, the two-temperature conditions are ineffectual.
Increasing $\epsilon\subs{}$ decreases the frequency spacing of the modes.
Varying $\epsilon\subs{}$ alters the instability of each mode
at a rate that depends on the harmonic number $n$,
and the two-temperature flow parameters $(\sigma\subs{},\psiei)$.
Except when $\epsilon\subs{}$ is very large
and $\sigma\subs{}$ is small,
the $\dR$ of each mode only ever decreases
with an increase in $\epsilon\subs{}$.
\item
The effect of varying the electron-ion coupling $\psiei$ is
insignificant in the bremsstrahlung-cooling dominated regime 
($\epsilon\subs{}$ small).
Decreasing $\psiei$, making the electron-ion exchange inefficient,
tends to destabilize the accretion column in all modes,
and increases the frequency spacing of the modes slightly.
\item
Varying the ratio of electron and ion pressures at the shock
$\sigma\subs{}$
affects the frequency spacing of the modes in non-simple ways.
For some choices of the global parameters a given change in $\sigma\subs{}$
increases the frequency spacing;
in other cases it causes a decrease.
Decreasing $\sigma\subs{}$ usually reduces the instability of modes,
and this is accentuated when $\epsilon\subs{}$ is high.
\item
Transverse perturbations yield $\dR$ maxima in $\kappa$
that can destabilize modes that are stable to purely longitudinal oscillations.
If $\sigma\subs{}$ is not too small
then these maxima may enter the stable regime again
for sufficiently great $\epsilon\subs{}$.
\end{enumerate}

\section{Discussion}
\label{'discussion'}

Most previous perturbative analyses of radiative shocks
(\eg Chevalier \& Imamura 1982, Bertschinger 1986, 
Houck \& Chevalier 1992,  T\'{o}th \& Draine 1993, Dgani \& Soker 1994) 
represented the cooling function as a single-power-law term,
and set the indices to various values
to mimic the presence of different physical cooling processes.
Such formulations do not adequately describe systems
where more than one cooling process is significant,
especially where these processes
differ greatly in their innate influences over thermal instability.
The realistic interplay between the influences of
a stabilizing cooling process and a destabilizing cooling process
was explicitly included in
Saxton \etal (1997) and Saxton \etal (1998).
This condition is retained in the present extended treatment,
and by taking the one-temperature limit
we recover the earlier results.

Our investigation considers the additional effects of
unequal electron and ion temperatures,
as in the work of Imamura \etal (1996),
which considered accretion onto non-magnetic white dwarfs.
Ours combines
the two-temperature treatment with
the effects of the joint presence of bremsstrahlung and cyclotron cooling.
Therefore our analysis is applicable to
radiative accretion shocks on white dwarfs in 
magnetic cataclysmic variables with
radiative cooling timescales possibly comparable to or faster than
the electron-ion energy exchange.

The accretion column in a mCV
is not geometrically extended in 
the vertical direction (see \eg Cropper 1990),
and the field-aligned flow suffers no transverse mass flux
(unlike the case of Dgani \& Soker 1994).
Our the cases of nonzero $\epsilon\subs{}$ correspond to magnetic fields
much stronger than those investigated by
Hujeirat \& Papaloizou \shortcite{hujeirat},
and therefore the field can not develop large transverse kinks
that would suppress oscillations by the mechanism of
T\'{o}th \& Draine \shortcite{toth}.
The altitude of the shock above the white dwarf surface 
is negligible compared to the size of the white dwarf,
so the variation of gravity in the accretion column,
studied by Houck \& Chevalier (1992), is unimportant.
Therefore we do not investigate effects of non-planar geometry
(as in Bertschinger 1986). Transverse perturbations,
which are manifest as corrugation of the shock structure and oscillation,
are studied, (as in Imamura \etal 1996).

Except when the energy exchange of electrons and ions is very weak,
increasing the efficiency of the cyclotron cooling $\epsilon\subs{}$
(i.e. strengthening the ambient magnetic field)
stabilizes each mode monotonically.
The modes stabilize at independent rates in $\epsilon\subs{}$,
meaning that some of those modes which are highly unstable
in the weak-field regime
may actually stabilize at low thresholds of $\epsilon\subs{}$.
The detailed behaviour of the mode stabilization in $\epsilon\subs{}$
depends on the parameters $\sigma\subs{}$ and $\psiei$.

As in Saxton \etal (1998) the instability of a particular mode
does not imply the instability of all higher modes,
when more than a single cooling process is significant.
This characteristic of competition between cooling processes
persists in the general two-temperature and non-longitudinal oscillation cases.
If cyclotron cooling is significant relative to bremsstrahlung cooling,
the electron-ion exchange process determines
the sequence of $\dR$,
but the frequency spacing of the quantized $\dI$ sequence
depends mainly on $\epsilon\subs{}$.
Cases of bremsstrahlung-cooling dominated shocks are similar
and their properties are scarcely affected by two-temperature effects.

The basic sequence of mode stabilities in the bremsstrahlung-cooling 
dominated case
has a stable fundamental and each higher mode less stable than its predecessor.
Strengthening cyclotron cooling tends to stabilize all the modes,
but does so in a fashion that is individual to each mode,
so that deviations from the trend develop as $\epsilon\subs{}$ increases
(especially in the presence of two-temperature effects).
For low $\epsilon\subs{}$ these stability deviations are slight.
For greater $\epsilon\subs{}$,
the $\dR$ sequence of the modes is a less simple pattern
depending on the parameters of the shock pressures $\sigma\subs{}$
and electron-ion exchange $\psiei$.

Two-temperature effects dominate the properties of the modes
when $\epsilon\subs{}$ is very high,
and beyond some threshold in $\epsilon\subs{}$,
which depends on $(\sigma\subs{},\psiei)$,
two-temperature effects have a destabilizing influence.
If the electron-ion exchange is sufficiently weak,
the exchange process dominates the oscillations
and some modes that would be stable at lower $\epsilon\subs{}$
become unstable again at higher $\epsilon\subs{}$.

The non-simple order in which the modes are stabilized
means that the thermal instability properties of accretion 
shocks in mCVs are not simply a trivial result of a relationship 
between oscillatory, cooling and energy exchange timescales.
The competing effects must be considered in detail. However we 
note that our analysis proves modes to be unstable, but does not 
necessarily prove stability.
Minorsky (1962, chapter 14) uses topological and perturbative 
analytic arguments to consider the general conditions of stability
of non-linear oscillatory systems. This issue was also addressed 
by Lyapunov in his analysis of stability of dynamical systems 
(Lyapunov 1966, see section 1.5.3, Leipholz 1970 for another 
treatment of Lyapunov's theorems on stability). In summary, 
the linear terms dominate in the limit of small amplitudes, i.e.\  
every mode that is linearly unstable remains unstable in a full 
non-linear treatment, although higher-order contributions may 
cause a stable limit cycle at larger amplitudes.

\section{Conclusions}  
\label{'conclusions'}

We investigated plane-parallel radiative shocks
in which bremsstrahlung and cyclotron cooling
may be fast compared with electron-ion energy exchange.
Parameters and boundary conditions are chosen to be
appropriate for accretion onto the surface of a magnetic white dwarf.
The cooling function was an explicit sum of contributions from
thermal bremsstrahlung (which destabilizes the flow)
and cyclotron cooling (which has a stabilizing tendency).

The relative efficiency of the cyclotron emission
$\epsilon_{\rm s}$ is varied. Stability of the shock to longitudinal 
and non-longitudinal perturbations was investigated. The ratio of electron 
and ion pressures at the shock $\sigma\subs{}$, and the rate of 
electron-ion energy exchange $\psiei$ were also varied.
Variations of the exchange, pressure and cooling parameters,
$(\sigma\subs{},\psiei,\epsilon\subs{})$,
affect the stability and frequency of each mode in a manner
that is individual to the mode and the conditions of the stream.
Increasing $\epsilon\subs{}$ increases the stability of each mode,
until a threshold value in $\epsilon\subs{}$ is exceeded.
Beyond this point, increasing $\epsilon\subs{}$ destabilizes a mode slowly.
This threshold is very high except when $\psiei$ is very small.

The introduction of unequal electron and ion temperatures has little 
effect on the modes of a bremsstrahlung-cooling dominated accretion column. 
In these cases, the higher harmonics tend to be less stable to oscillations.
When bremsstrahlung and cyclotron cooling are comparable in strength,
the pattern of the modes' stabilities becomes less regular,
and two-temperature effects tend to stabilize the shock.
For a cyclotron-cooling dominated shock all of the modes are stable
in the single-temperature limit,
but introduction of two-temperature effects can make them less stable.
In some cases of small $\psiei$ one or more of the lower modes
may be unstable even for high $\epsilon\subs{}$.
Unlike the bremsstrahlung emission,
cyclotron cooling is strongest near the shock,
where the difference of electron and ion temperatures is greatest.
Therefore the slightest presence of two-temperature effects
means that the stability of a cyclotron-cooling dominated shock
is dominated by the energy exchange between the two fluid components
rather than the cooling function.

Many cases exist where the mode $n$ is unstable 
while the mode $n+1$ is stable, and some higher modes are unstable.
This is a general characteristic of radiative shocks
with competing stabilizing and destabilizing cooling processes,
and is preserved in the generalization to conditions with
two-temperatures and non-longitudinal perturbations.

The oscillatory parts of the dimensionless eigenfrequencies $\dI$
follow a sequence 
which is approximately regular and linear.
In the single-temperature case this sequence resembles
the modes of a pipe open at one end:
$\dI\approx\dIO(n-1/2)+\dC$.
When two-temperature effects are present
the limit of large $\epsilon\subs{}$ causes the frequencies
to approach quantization like a doubly-open pipe
$\dI\approx\dIO n$,
because of the fixed velocity condition at the white dwarf surface
becomes less significant than the electron-ion thermal disparity
throughout the column.
The frequency spacing of the modes, $\dIO$,
always decreases as $\epsilon\subs{}$ increases,
but tends to increase when the electron-ion exchange is weaker.
The frequency offset 
$\dC$ is small for low $\epsilon\subs{}$ and 
it generally increases with increasing $\epsilon\subs{}$;
in the limit of cyclotron cooling and two-temperature dominance
the offset approaches $\dC\rightarrow{\frac12}\dIO$,
giving the doubly-open pipe behaviour.

Introducing transverse perturbations can destabilize modes that are stable
to purely longitudinal perturbations,
except when cyclotron cooling dominates
or two-temperature effects are strong.
The modes become non-unique in $\dI$, for different $\kappa$.
In the limit of large $\kappa$ all modes are stable
with oscillatory parts quantized like a doubly-open pipe $\dI\propto n$,
regardless of the manner of their quantization
under the same parameters in the absence of transverse perturbations.
Each mode has a number of instability maxima in $\kappa$,
and these maxima evolve
as the parameters $(\sigma\subs{},\psiei,\epsilon\subs{})$ change.
When the pressure ratio at the shock $\sigma\subs{}$ is small,
the maxima of $\dR$ are stabilized at lower $\epsilon\subs{}$.
When the electron-ion exchange is weak ($\psiei$ is small),
the maxima may continue to destabilize at high $\epsilon\subs{}$.

\section{ACKNOWLEDGEMENTS}

KW acknowledges the support from the Australian Research Council
through an ARC Australian Fellowship.


\begin{thebibliography}{99}

\bibitem[\protect\citename{Aizu }1973]{aizu}
Aizu, K., 1973. Prog. Theor. Phys., 49, 1184  
\bibitem[\protect\citename{Bertschinger }1986]{bertschinger}
Bertschinger, E., 1986, ApJ, 304, 154
\bibitem[\protect\citename{Chanmugam et al.\ }1985]{cls}
Chanmugam G., Langer, S. H., Shaviv, G., 1985, ApJ, 299, L87 
\bibitem[\protect\citename{Chevalier \& Imamura }1982]{chevalier}
Chevalier, R. A., Imamura, J. N., 1982, ApJ, 261, 543 
\bibitem[\protect\citename{Cropper }1990]{cropper}
Cropper, M., 1990, Sp. Sci. Rev., 54, 195
\bibitem[\protect\citename{Cropper, Wu, Ramsay \& Kocabiyik }1999]{cropper99}
Cropper, M., Wu, K., Ramsay, G. Kocabiyik, A., 1999, MNRAS, 306, 684
\bibitem[\protect\citename{Dgani \& Soker}1994]{dgani}
Dgani, R., Soker, N., 1994, ApJ, 434, 262
\bibitem[\protect\citename{Gaetz, Edgar \& Chevalier }1988]{gaetz}
Gaetz, T. J., Edgar, R. J., Chevalier, R. A., 1988, 
    ApJ, 329, 927 
\bibitem[\protect\citename{Falle }1975]{falle75}
Falle, S. A. E. G, 1975, MNRAS, 172, 55
\bibitem[\protect\citename{Falle }1981]{falle81}
Falle, S. A. E. G, 1981, MNRAS, 195, 1011
\bibitem[\protect\citename{Houck \& Chevalier }1992]{houck}
Houck, J. C., Chevalier, R. A., 1992, ApJ, 395, 592 
\bibitem[\protect\citename{Hujeirat \& Papaloizou }1998]{hujeirat}
Hujeirat, A., \& Papaloizou, J. C. B., 1988, A\&A, 340, 593
\bibitem[\protect\citename{Imamura }1985]{imamura85}
Imamura, J. N., 1985, ApJ, 296, 128   
\bibitem[\protect\citename{Imamura \& Wolff }1990]{wolff90}
Imamura, J. N., Wolff, M. T., 1990, ApJ, 355, 216   
\bibitem[\protect\citename{Imamura, Wolff \& Durisen }1984]{durisen}
Imamura, J. N., Wolff, M. T., Durisen, R. H., 1984, 
    ApJ, 276, 667 
\bibitem[\protect\citename{Imamura et al.\ }1996]{aboasha}
Imamura, J. N., Aboasha, A., Wolff, M. T., Wood, K. S., 
    1996, ApJ, 458, 327 
\bibitem[\protect\citename{Innes, Giddings \& Falle }1987a]{innes87a}
Innes, D. E., Giddings, J. R., Falle, S. A. E. G., 1987a,  
    MNRAS, 224, 179  
\bibitem[\protect\citename{Innes, Giddings \& Falle }1987b]{innes87b}
Innes, D. E., Giddings, J. R., Falle, S. A. E. G., 1987b,  
    MNRAS, 226, 67   
\bibitem[\protect\citename{King \& Lasota }1979]{king}
King, A. R., Lasota, J. P., 1979, MNRAS, 188, 653 
\bibitem[\protect\citename{Lamb \& Masters }1979]{lamb}
Lamb, D. Q., Masters, A. R., 1979, ApJ, 234, L117  
\bibitem[\protect\citename{Langer, Chanmugam \& Shaviv }1981]{langer81}
Langer, S. H., Chanmugam, G., Shaviv, G., 1981, ApJ, 245, L23 
\bibitem[\protect\citename{Langer, Chanmugam \& Shaviv }1982]{langer82}
Langer, S. H., Chanmugam, G., Shaviv, G., 1982, ApJ, 258, 289 
\bibitem[\protect\citename{Leipholz }1970]{leipholz}
Leipholz, H., 1970, Stability Theory: An Introduction to the Stability of Dynamic Systems and Rigid Bodies, Academic Press, New York 
\bibitem[\protect\citename{Lyapunov }1966]{lyapunov}
Lyapunov, A. M., 1966, Stability of Motion, Academic Press, New York
\bibitem[\protect\citename{Melrose }1986]{melrose}
Melrose, D. B., 1986, Instabilities in Laboratory and Space Plasmas,
Cambridge University Press, Cambridge
\bibitem[\protect\citename{Minorsky }1962]{minorsky}
Minorsky, N., 1962, Nonlinear Oscillations, Van Nostrand, Princeton
\bibitem[\protect\citename{Rybicki \& Lightman }1979]{rybicki}
Rybicki, G. B., Lightman, A. P., 1979, Radiative Processes 
    in Astrophysics, John Wiley \& Sons, New York
\bibitem[\protect\citename{Saxton }1999]{saxtonthesis}
Saxton, C. J., 1999, PhD Thesis, University of Sydney, Australia.
\bibitem[\protect\citename{Saxton, Wu \& Pongracic }1997]{saxton97}
Saxton, C. J., Wu, K., Pongracic, H., 1997, 
    Publ. Astr. Soc. Australia, 14, 164 
\bibitem[\protect\citename{Saxton, Wu, Pongracic \& Shaviv}1998]{saxton98}
Saxton, C. J., Wu, K., Pongracic, H., Shaviv, G., 1998, 
    MNRAS, 299, 862
\bibitem[\protect\citename{T\'{o}th \& Draine}1993]{toth}
T\'{o}th, G., Draine, B. T., 1993, ApJ, 413, 176  
\bibitem[\protect\citename{Warner }1995]{warner}
Warner, B., 1995, Cataclysmic Variable Stars.
    Cambridge Univ. Press, Cambridge, p.256
\bibitem[\protect\citename{Wolff, Gardiner \& Wood }1989]{wolff89}
Wolff, M. T., Gardner, J., Wood, K. S., 1989, ApJ, 346, 833 
\bibitem[\protect\citename{Wu }1994]{wu}
Wu, K., 1994, Proc. Astr. Soc. Australia, 11, 61 
\bibitem[\protect\citename{Wu, Chanmugam \& Shaviv }1992]{wcs92}
Wu, K., Chanmugam, G., Shaviv, G., 1992, ApJ, 397, 232 
\bibitem[\protect\citename{Wu, Chanmugam \& Shaviv }1994]{wcs94}
Wu, K., Chanmugam, G., Shaviv, G., 1994, ApJ, 426, 664
\bibitem[\protect\citename{Wu et al.\ }1996]{wpcs}
Wu, K., Pongracic, H., Chanmugam, G., Shaviv, G., 1996, 
    Publ. Astron. Soc. Aust., 13, 93

\end{thebibliography}
\end{document}